\documentclass[epjST]{svjour}
\pdfoutput=1

\usepackage[T1]{fontenc} 
\usepackage[utf8]{inputenc}
\usepackage[english]{babel}
\usepackage{lmodern}
\usepackage{url}

\usepackage{amsmath,amssymb}
\usepackage[numbers]{natbib}
\usepackage{bm}
\DeclareSymbolFont{usualmathcal}{OMS}{cmsy}{m}{n}
\DeclareSymbolFontAlphabet{\mathcal}{usualmathcal}

\usepackage{graphicx}
\usepackage{xcolor} 
\DeclareGraphicsExtensions{.pdf,.png,.jpg} 
\usepackage{subfigure} 

\def\rho{\varrho}
\def\ge{\geqslant}
\def\geq{\geqslant}

\let\eps\varepsilon
\let\rho\varrho
\let\theta\vartheta
\def\ignore#1{}

\begin{document}
\title{A Random Growth Model for Power Grids and Other Spatially Embedded Infrastructure Networks}
\author{Paul Schultz\inst{1,2}\fnmsep\thanks{\email{pschultz@pik-potsdam.de}} \and
Jobst Heitzig\inst{1} \and J\"urgen Kurths\inst{1,2,3} }
\institute{Potsdam Institute for Climate Impact Research, 14412 Potsdam, Germany \and
 Department of Physics, Humboldt University Berlin, 12489 Berlin, Germany \and  
 Institute for Complex Systems and Mathematical Biology, University of Aberdeen, AB24 3UE Aberdeen, UK}
\abstract{
We propose a model to create synthetic networks that may also serve as a narrative of a certain kind of infrastructure network evolution. It consists of an initialization phase with the network extending tree-like for minimum cost and a growth phase with an attachment rule giving a trade-off between cost-optimization and redundancy. Furthermore, we implement the feature of some lines being split during the grid's evolution. We show that the resulting degree distribution has an exponential tail and may show a maximum at degree two, suitable to observations of real-world power grid networks. In particular, the mean degree and the slope of the exponential decay can be controlled in partial independence. To verify to which extent the degree distribution is described by our analytic form, we conduct statistical tests, showing that the hypothesis of an exponential tail is well-accepted for our model data. 
\keywords{power grid -- numerical simulation -- network topology -- random growth model -- degree distribution -- minimum spanning tree -- redundancy -- construction costs}
} 

\maketitle

\begin{center}
The final publication is available at \url{link.springer.com} via the following DOI:
 \url{http://dx.doi.org/10.1140/epjst/e2014-02279-6}.
\end{center}

\section{Introduction}

\subsection{Motivation}

Currently, the German high-voltage transmission grid alone has an astonishing overall line length of about 150,000\ km
\cite{BDEW2013}. 
While this is a well-known fact, the exact topology of the power grid network remains generally unavailable or confidential, 
making it difficult to find suitable data sources for power grid research. 
Nevertheless, this is a fruitful field of research with increasing importance; confronting challenges such as 
decentralized generation (in the low-voltage grid), 
an increased frequency and severity of extreme  events due to climate change 
and the construction of long-ranging high-voltage interconnections to cope with power supply and demand being spatially more separated. 
Many of the challenges are of a topological nature, 
e.g. related to unstable structures and their improvement \cite{Menck2012} or to ``paradoxical'' overloads \cite{Witthaut2012}.
To our knowledge, there are only a few real-world power grid data sets available for research,
e.g., the publicly available high-voltage grid from National Grid UK\footnote{\url{http://www2.nationalgrid.com/uk/services/land-and-development/planning-authority/} (accessed \today)}
seems to be the only accessible data set of a contemporary power grid that is geo-referenced, 
i.e., comes with data on geographic node locations. 
Further commonly used data sets are the IEEE test cases with networks ranging from about ten to 300 nodes\footnote{\url{http://www.ee.washington.edu/research/pstca/} (accessed \today)}
or the power grid\footnote{\url{http://www-personal.umich.edu/~mejn/netdata/}(accessed \today)} of the Western United States \cite{Watts1998}.
Many research questions however require to a certain extend numerical simulations
and thus there is a need for easy-to-implement models producing ensembles of synthetic though ``realistic'' power grid topologies, 
tunable to observed statistical properties. 

As power grids display some topological similarities to some other infrastructure networks, 
in particular an approximately exponentially decaying degree distribution,
our aim here is to construct a random graph model that is well-suited for the power grid use-case
but is also tunable to other use-cases with exponential decay,
e.g., railroad \cite{Sen2003} or road \cite{Rosvall2005} networks, 
or certain information infrastructure networks \cite{Dorogovtsev2002}. 
Note that it is often difficult to distinguish from data whether a degree distribution of a network with only several hundreds of nodes
follows a power law (as it is often claimed by presenting a double-logarithmic plot that seems to be more or less linear) 
or some other heavy-tailed distribution, e.g., an exponential;
hence random grid models with exponential decay may have an even broader use than it seems at first glance.

Following the above demands, we propose here a model to create synthetic networks, resembling key features of real-world power grids
concerning the degree distribution and other common statistical properties from complex networks analysis as well. 

We will show that our proposed model offers features to produce a wide range of 
networks with an exponentially decaying degree distribution 
that arises endogenously from some simple and plausible construction and growth mechanisms
involving linking criteria from the real world such as the minimization of distances,
the maximization of redundancy, or the splitting of links. 

\subsection{Topological properties of real-world power grids}

\begin{figure}[htp]
\centering
\includegraphics[width=.9\textwidth]{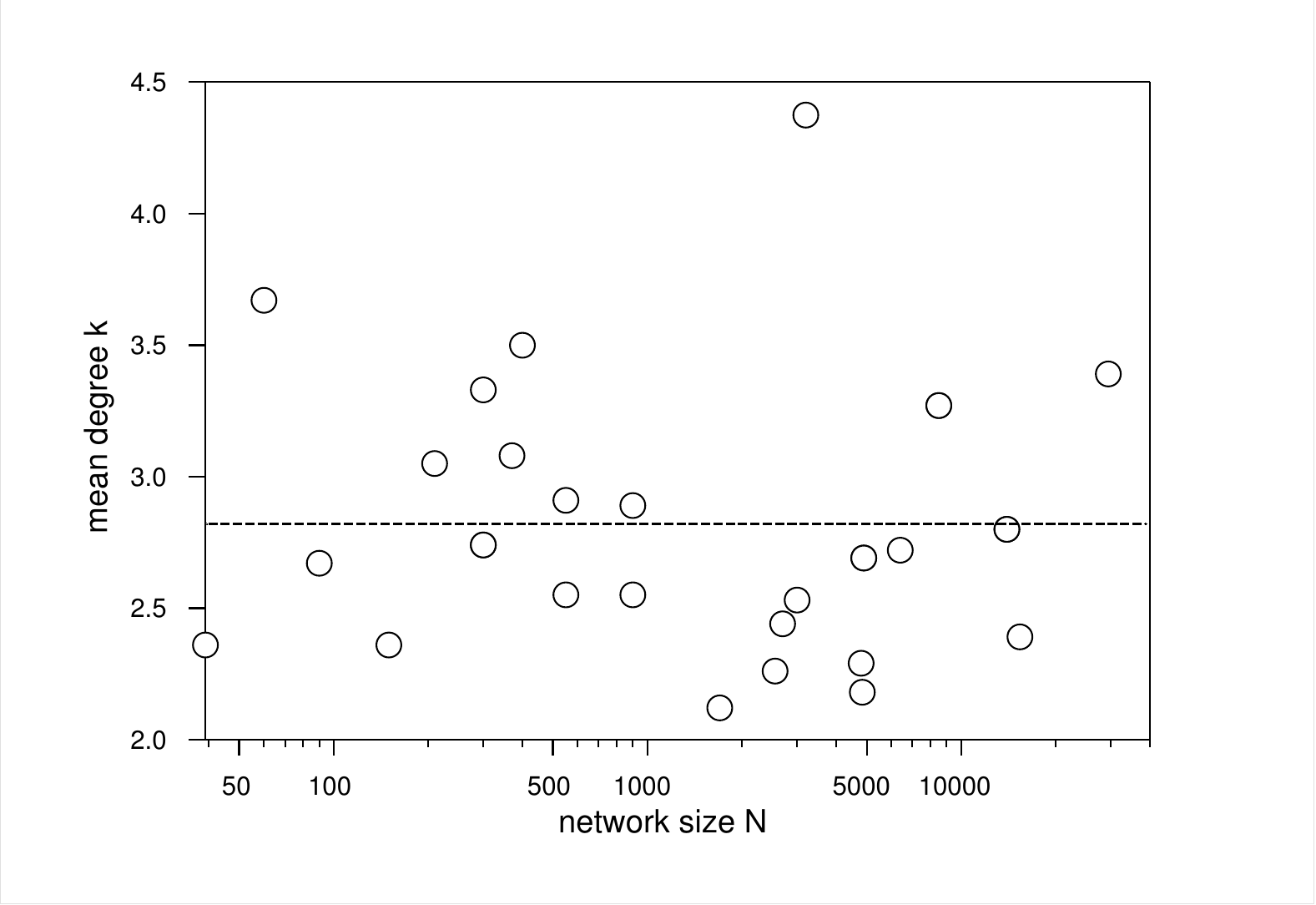}
\caption{Semi-logarithmic plot of the mean degree observed for various power grids in different areas \cite{Pagani2013} versus the corresponding system size (number of nodes in the network). The dashed line indicates the mean value of $k=2.8$.}
\label{fig:k_vs_N}
\end{figure}

In the following we intend to summarize topological features of power grid networks. 
We assume that a power grid is well-described by a two-dimensional, spatially embedded, connected network. 
To fit into the most common complex networks theory framework, 
we treat the network as undirected, unweighted and simple, meaning no two lines connecting the same endpoints (no multiple edges). 
In reality however, multiple edges indeed occur regularly as multiple circuits hosted in parallel are on the same towers (see Sec.\ \ref{sec:discussion}).
Note that despite their two-dimensional spatial embedding, 
power grids usually have crossing lines (that appear to ``intersect'' in two-dimensional plots),
so power grids are not ``planar'' in the graph-theoretical sense.

Data sets studied in the literature represent networks with a size $N$ ranging from dozens to thousands of nodes, 
and they exhibit an {\em extreme sparsity} with a {\em mean degree} $\bar k\approx 2.8$ \cite{Pagani2013} that is independent of $N$ (see Fig.\ \ref{fig:k_vs_N}).

Another remarkable feature of real-world power grids is that their {\em degree distribution,} i.e.,
the statistical distribution $p_k$ of the number of neighbors $k$ of a node,
appears to have a local maximum at very small degrees (see the slight ``bump'' at $k=2,3$ in Fig.\ \ref{fig:power}) \cite{Pagani2013,Scaglione2010}
and an {\em exponentially decaying tail} \cite{Barthelemy2011,Pagani2013,Rosas-Casals2009} of the form $p_k\sim e^{-k/\gamma}$.
The decay parameter $\gamma$ is typically estimated to be between $1.5$ and $2$ \cite{Pagani2013,Rosas-Casals2009}. 
However, it often remains unclear in the literature how the hypothesis of an exponential decay was tested against possible alternative heavy-tailed distributions. 

As the observed kind of degree distributions could not be explained by tree-like networks 
--- they have exactly $N-1$ edges and thus a mean degree of
$2(N-1)/N \thicksim 2$ and no large individual degrees --- 
this hints at the existence of heterogeneously distributed \emph{redundancy}. 
The so-called ``$N-1$ criterion'' applied in power grid planning ensures that all nodes are connected to the grid via multiple circuits, 
which may however be hosted on the same towers.
Thus a network representation of the power grid that includes multiple lines would be at least 
{\em twice connected} in the graph-theoretical sense of topological connectedness, 
meaning that the removal of a single line can never split the grid into two disconnected parts.
Still, most network representations of power grids represent multiple parallel lines by only a single network link,
hence they are typically only {\em simply connected,}
meaning that there is always a link (which may represent multiple parallel circuits hosted on the same towers) 
whose removal would cause a separation of the network into two disconnected parts (though one of the parts would usually be very small); 
typical examples of such non-redundant links are tree-like appendices,
and indeed there is evidence that such structures may cause certain types of instability (``dead ends'' and ``dead trees'', \cite{Menck2014}). 
Despite this formally low level of topological connectedness, most links of a power grid are typically redundant.
E.g., in the example of Fig.\ \ref{fig:power}, the network has $1.3$ times as many edges as a tree of same size would have.
 
Another metric for redundancy is {\em algebraic connectivity} $\lambda_2$ (also called the Fiedler eigenvalue), 
which is the second-smallest eigenvalue of the network's Laplacian matrix, with the smallest eigenvalue being zero. 
It is also related to the minimum number of links $R$ that would disconnect the network in two parts and lies between zero (unconnected network) 
and the value of the topological connectivity (which is one for power grids, see above).
For a fixed network size $N$, the algebraic connectivity increases linearly with $R$, $\lambda_2\propto R$, meaning that additional redundancy 
relates to higher values of $\lambda_2$. On the contrary, for fixed $R$, the algebraic connectivity decreases with $N$ as $\lambda_2\propto N^{-1}$.
This scaling behavior is well visible for the data points in  Fig.\ \ref{fig:lambda2_vs_N}, see also the discussion in Sec.\ \ref{subsec:monte-carlo}.

Finally, also the network's {\em clustering coefficient,} i.e., the probability that two neighbors of a node are also directly linked,
can be seen as a metric for redundancy, though a very local one. Empirically there seems to be no dependence on the size of the network (see Tab.\ \ref{tab:ucte}), what is also the case for realizations of our model (see discussion below).

Much of the redundancy in a power grid is indeed relatively local,
which can also be seen from the fact that the network {\em average shortest path length (aspl),} i.e., 
the average number of links required to pass from any node to another, grows with $N$ at a rate somewhere between 
$O(\sqrt[5]N)$ and $O(\sqrt N)$ (the latter being the aspl of a two-dimensional regular grid, see Fig.\ \ref{fig:aspl_vs_N_INSET_DATA}).

Power grids are sometimes thought to possess the {\em small-world property,} 
i.e., having a low {\em average shortest path length} but a high clustering coefficient compared to a random graph of same size \cite{Pagani2013}. 
Still, the predominant model for small-world networks, the Watts-Strogatz (WS) model \cite{Watts1998}
has other statistical properties that are very different from power grids and can thus not serve as a model for power grids.
In particular, WS small-world networks have a mean degree that is well above $\log N$ 
which is however not at all the case for power-grids (see above).

Regarding the growth of power grids, 
there is evidence for two phases of power grid construction, an initialization and an accelerated growth phase, 
as it was found for the development of the French high-voltage grid \cite{Rosas-Casals2009}. 
There the network develops initially coarsely and tree-like to quickly cover the spatial extents with minimal costs (line length). 
This is followed by a densification of the network (the appearance of many short-ranged links) and creation of additional redundancy. 
It can also be observed that lines may split in two by addition of a new node. 

\subsection{Existing network models}

The network models that are probably most popular in the literature 
are random graphs based on the algorithms of Erd\H{o}s-R\'{e}nyi \cite{Erdos1959} or Gilbert \cite{Gilbert1959}, 
the small-world model from Watts and Strogatz \cite{Watts1998}, 
and finally the scale-free networks based on Barab\'{a}si and Albert's preferential attachment \cite{Barabasi1999}. 

Random graphs are typically not sparse, have an aspl well below that of power grids, and show a non-heavy-tailed, binomial degree distribution. 

The simplest approach to create spatially embedded networks are so-called {\em random geometric networks} \cite{Herrmann2003}
where node locations are drawn randomly from the unit square and nodes $i,j$ are linked iff their distance $d_{ij}$ is below a threshold $\eps$.
However, such graphs are not necessarily connected, their mean degree grows as $\approx 4N\eps^2$ instead of being constant,
their degree distribution is Poissonian, thus decaying faster than exponentially,
and both their clustering coefficient and aspl are larger than those of power grids 
($\approx (3/4)^2\approx 0.57$ and $\approx \sqrt{9N/4\bar k}$, respectively).
A model with similar properties is the Waxman model \cite{Waxman1988}, where two nodes are linked with a probability proportional to $\exp{-d_{ij}/\eps}$. 

Scale-free networks on the other hand have much more heavy-tailed, power-law degree distributions,
and their mean degree is not continuously adjustable. 
Barab\'{a}si and Albert show however that, in the case of equal (non-preferential) connection probabilities, 
their growth model produces exponentially decaying degree distributions. 
This is reinforced in \cite{Dorogovtsev2002} who show that the corresponding master equation has the stationary solution $p_k=2^{-k}$. 
A similar result is found in \cite{Callaway2001}, where the authors present a growth model that is again not spatially embedded, 
but shows an exponentially decaying degree distribution. 
This indicates that a model for power grid topologies can be designed as a growth model that results in an observably exponentially decaying degree distribution.

The Watts-Strogatz small-world algorithm \cite{Watts1998} with its rewiring parameter $p$ is often used in studies that focus on the transition from regular to random topologies, 
but it is not suited to generate power grid topologies. 
Its mean degree is not continuously adjustable, it has a too small aspl (for large $p$) or is too regular (for small $p$) 
and again is not a growth model and has no spatial embedding.

Any desired degree distribution, including an exponentially decaying one, 
can be generated using the so-called {\em configuration model} \cite{Molloy1995a,Newman2001}
in which node degrees are prescribed endogenously for all nodes and the links are generated respecting these degrees but in an otherwise random manner.
Still, this does not ensure the correct behavior of other statistical network measures
and the construction mechanism does not seem a plausible assumption for the case of power grids
where the exponential decay is very likely not prescribed by design but rather emerges endogenously.


The ``RT-nested-Smallworld''-model \cite{Scaglione2010} is to our knowledge the first attempt to design a model 
that especially matches the statistical properties of power grid topologies, 
by combining and modifying standard components of existing network models. 
Their algorithm combines a modified small-world model for generating local structures which are then joined in a relatively regular mesh-like way.
Despite its matching several statistical properties of real-world networks, 
it hence seems to produce networks that are a little too regular, 
although the authors unfortunately provide no plots of these networks and some of the actual construction details remain somewhat unclear.
In addition, their model is not spatially embedded, 
hence their construction of links is in no way based on the locations of nodes, 
despite the fact that node locations are certainly an important criterion for actual power transmission line construction. 

Although it has not yet been used in random network models to our knowledge, 
we finally have to mention the concept of a {\em minimum spanning tree (MST)} \cite{Boruvka1926,Boruvka1926a}, 
An MST connects a given set of nodes in a tree-like manner with those $N-1$ many lines that minimize the sum of some edge weights 
(i.e. the spatial distances between the nodes);
there are efficient algorithms to solve this minimization problem \cite{Kruskal1956,Prim1957}. 
Interestingly, although \cite{Boruvka1926,Boruvka1926a} invented MSTs to design the Moravian power grid, 
they apparently have not been considered in the academic power grid literature later on.
A model based only on MSTs, e.g., constructing the MST of a random set of locations in the plane,
would of course be much too sparse ($\bar k=2(N-1)/N\approx 2$) since trees have no redundancy, 
would have a flat degree distribution (no degrees above five),
would produce no line crossings, and would not consider network growth.
Still, because of the obvious intuitive appeal of the cost-minimization criterion,
the model we present below will use MSTs as an important building block.

\subsection{Goal and approach}

To support theoretical and numerical research on the influence of network topology, 
on the efficiency and stability of power grids and other spatially embedded infrastructure networks, 
we aim at constructing a random network growth model that has the following main features. It\ldots
\begin{description}
  \item[i)] has any number of {\em spatially embedded} nodes and links.
  \item[ii)] supports both {\em random} and {\em exogeneously given} node placement.
  \item[ii)] has an {\em initialization phase} and a {\em growth phase}.
  \item[iv)] is {\em connected} but {\em sparse} with a {\em tunable mean degree}.
  \item[v)] has a {\em tunable trade-off between cost minimization and redundancy}.
  \item[vi)] is easy to implement and of little computational complexity to {\em support Monte-Carlo simulations}.
  \item[vii)] is based on an idealized but {\em plausible construction mechanism}.
  \item[vii)] has an {\em exponentially} decaying degree distribution
  		that is not imposed exogeneously but emerges endogenously from the growth 		
  		process, with a {\em tunable decay rate}.
\end{description}

Our approach mainly focuses on minimizing costs 
via a minimal spanning tree in the initialization phase and via links to nearest existing nodes in the growth phase. 
The initialization phase results in an initially coarse, tree-like grid to quickly construct a far-reaching supply network with minimal costs. 
In parallel we raise redundancy via additional links that optimize a simple heuristic function estimating the trade-off between cost and redundancy in the growth phase as well.

\section{The model}

\subsection{Parameters and definitions}

\paragraph{Redundancy/cost optimization.}

Our model uses the following heuristic target function for redundancy/cost optimization when adding individual links:
\begin{align}
  f(i,j,G) &= \frac{(d_G(i,j)+1)^r}{d_{\rm spatial}(x_i,x_j)} \label{eq:cost_function}
\end{align}
where $d_G(i,j)$ is the length of a shortest path between nodes $i,j$ in network $G$ (i.e., their network distance measured in hops).

The rationale for using this heuristic target function $f$ 
instead of a target function directly based on a more complex redundancy metric such as $\lambda_2$ is the following. 
We are modeling very sparse connected networks, 
which are hence almost tree-like and contain many tree-shaped appendices.
Call a link in a network {\em redundant} iff the network remains connected when that link is removed. 
If a link $i$---$j$ is added to a tree $G$, 
then exactly $d_G(i,j)+1$ many links of the resulting network are redundant (since they form a circle).
If $i$---$j$ is added to an almost tree-like $G$, at least $d_G(i,j)+1$ many (and not too many more) links become redundant.
The notion of redundant link fits better with security rules actually used by grid operators 
than more complex metrics from complex network theory such as $\lambda_2$ do.
Finally, the exponent $r$ lets us tune the relative importance given to redundancy vs. costs, 
from $r=0$ (redundancy is disregarded) to $r\to\infty$ (costs are disregarded).

\paragraph{Parameters} Our model has the following input parameters:
\begin{itemize}
  \item an initial number of nodes $\mathbf{N_0}\ge 1$,
  \item a probability $\mathbf{p}\in[0,1]$ for constructing an additional redundancy line attached at each new node,
  \item a probability $\mathbf{q}\in[0,1]$ for constructing a further redundancy line between existing nodes in each growth step,
  \item an exponent $\mathbf{r}\ge 0$ for the cost-vs-redundancy trade-off,
  \item a probability $\mathbf{s}\in [0,1]$ for splitting an existing line in each growth step,
  \item either a spatial probability distribution $\bm\rho$ (e.g. given by a
	probability density $\phi(x)$ such that $d\rho(x) = \phi(x)\,dx$) for the random placement of 
  	nodes (e.g. a uniform distribution in some two-dimensional region or a 
  	population density), or a sequence of exogeneously given node locations $x_1,x_2,
  	\dots$,
  \item a spatial distance function $\mathbf{d_{\rm spatial}(x,y)}$ representing the costs of 
  		building a line from $x$ to $y$ (e.g. Euclidean or geodesic distance, 
  	    or some measure of local per-unit cost integrated along a cheapest curve from 
  	    $x$ to $y$, etc.).
\end{itemize}
Mean degree and decay of the degree distribution 
are mainly determined by $p,q$ and $s$ (see \ref{subsec:analytical}),
while the trade-off between total line length and algebraic connectivity 
is mainly determined by $r$ (see \ref{subsec:monte-carlo}). \\\\
Next, we describe the network construction algorithm, which consists of two phases termed ``Initialization'' and ``Growth''.

\subsection{Initialization}

Given $N_0,p,q,r,s$, and $\rho$ or $x_1\dots x_N$, we construct the network $G$ as follows:
\begin{description}
  \item[I1.] If the locations $x_1\dots x_N$ are not given, draw them independently at random from $\rho$.
  \item[I2.] Initialize $G$ to be a minimum spanning tree (MST) for $x_1\dots x_N$ 
  		w.r.t.\ the distance function $d_{\rm spatial}(x,y)$ 
  		(using Kruskal's simple or Prim's more efficient algorithm \cite{Kruskal1956,Prim1957}).
  \item[I3.] Put $m = \lfloor N_0(1-s)(p+q)\rfloor$. 
  		For each $a = 1\dots m$, add a link to $G$ as follows:
  		Find that yet unlinked pair of distinct nodes $i,j\in\{1,\dots,N_0\}$ for which $f(i,j,G)$ (Eq.\ \ref{eq:cost_function}) is maximal,
  		and add the link $i$---$j$ to $G$.
\end{description}
Note that the resulting $G$ has $N_0-1+m\approx N_0(1+(1-s)(p+q))$ many links
(which is below the maximum possible number of $N_0(N_0-1)/2$ whenever $N_0\ge 7$, 
otherwise one has to adjust $m$ to $\min\{\lfloor N_0(1-s)(p+q)\rfloor,N_0(N_0-1)/2-(N_0-1)\}$).

\subsection{Growth}

Given $p,q,r$, a network $G$ of size $N$, and $\rho$ or a location $x_{N+1}$,
we add a new node $i=N+1$ to $G$ as follows.
With probability $1-s$, a node is added at a random position, 
is linked to its closest neighbor and maybe to a second node, 
and maybe an additional link is made between existing nodes (steps G1--G4 below);
otherwise (i.e., with probability $s$), a random link is split in two,
with a new node added halfway (step G5 below).
More precisely:
\begin{description}
  \item[G0.] With probabilities $1-s$ and $s$, perform either steps G1--G4 or step G5, respectively.
  \item[G1.] If $x_i$ is not given, draw it at random from $\rho$.
  \item[G2.] Find that node $j\in\{1,\dots,N\}$ for which $d_{\rm spatial}(x_i,x_j)$ is minimal
  		and add the link $i$---$j$ to $G$.
  \item[G3.] With probability $p$, 
  		find that node $\ell\in\{1,\dots,N\}\setminus\{j\}$ for which $f(i,\ell,G)$ is maximal,
  		and add the link $i$---$\ell$ to $G$.
  \item[G4.] With probability $q$, 
		draw a node $i'\in\{1,\dots,N\}$ uniformly at random,
  		find that node $\ell'\in\{1,\dots,N\}$ which is not yet linked to $i'$ 
  		and for which $f(i',\ell',G)$ is maximal,
  		and add the link $i'$---$\ell'$ to $G$.
  \item[G5.] Select an existing link $a$---$b$ uniformly at random,
		let $x_i=(x_a+x_b)/2$, 
		remove the link $a$---$b$, 
		and add two links $i$---$a$ and $i$---$b$.
\end{description}

\section{Results}
In the following section we present analytic properties, that can be derived from the above explained algorithm. Moreover,
we apply the model to illustrative examples and empirically derive further properties from a Monte-Carlo analysis.

\subsection{Analytic properties}\label{subsec:analytical}

Using the above algorithm, we let grow networks from an initial number $N_0$
of nodes up to some number $N\geq N_0$ of nodes and analyse their statistical
properties at that time point.

\subsubsection{Complexity}
The time complexity of the initialization is that of the algorithm used for finding $M$
(which is $O(N_0^2\log N_0)$ for Kruskal's algorithm \cite{Kruskal1956} and $O(N_0^2)$ for Prim's algorithm \cite{Prim1957})
plus $O(N_0^2m)$ in step I3, which makes a polynomial complexity of $O(N_0^3)$ in total.

If instead of the simple successive one-link-at-a-time redundancy/cost optimization in step I3,
one would seek the set of $m$ additional lines that optimizes some overall redundancy/cost trade-off
by an exhaustive search, this would likely have superexponential complexity of $O({N_0\choose m})=O((N_0/2)^{N_0})$,
and it is unclear whether a more efficient algorithm for this global optimization exists 
and is simple enough to serve as a plausible ingredient.

If instead of the heuristic function $f$ some target function $f'$ based on the resulting $\lambda_2$ would be used,
the complexity of calculating a single value $f'(i,j,G)$ would be at least that of calculating the eigenvalue $\lambda_2$ 
of an $N_0\times N_0$ matrix (at least $O(N_0^2)$, using for instance QR decomposition \cite{Kublanovskaya1962}),
which would lead to at least a complexity of $O(N_0^5)$ instead of $O(N_0^3)$ for step I3.

The computational complexity of adding a single node in the growth phase 
is dominated by the need to update the shortest path distance function $d_G$.
If $s=0$, this can be done in $O(N^2)$ time,
but if $s>0$ and a link is split in two, $d_G$ must be recalculated completely,
e.g. using Dijkstra's algorithm \cite{Dijkstra1959}, requires $O(N^2\log N)$ time.

In total, the generation of a network of size $N$ has thus 
complexity $O(N^3)$ for $s=0$ and $O(N^3\log N)$ for $s>0$.

\subsubsection{Degree distribution}
By construction, the  expected mean degree of $G$ is given by
\begin{align}
	\kappa &= 2(1+(1-s)(p+q)) - O(1/N) \approx 2 + 2p + 2q - 2ps - 2qs
\end{align}
at all times. It is bounded between $2$ ($p=q=0$ or $s=1$) and $6$ ($p=q=1$ and $s=0$).
Here, the term $2m/N_0=2(1-s)(p+q)$ is the contribution of links added in step I3.

\paragraph{Flat degree distribution of minimum spanning tree.}
For the generic two-dimensional case that we focus on here, 
all nodes in a MST have degrees between one and five 
since the angle between two straight spatially embedded links at the same node is always larger than $60^\circ$.
Numerical simulations reveal a unimodal distribution peaking at degree two 
and having almost no nodes with degree five (see Fig.\ \ref{fig:polars},a).  

\paragraph{Approximate distribution of large degrees in growth process.}
The approximate shape of the degree distribution for large $N-N_0$ can be estimated
using a similar approach as in \cite{Callaway2001} for the case of a not spatially embedded random growth model,
leading to an approximately exponential decay of the form $P(K=k)\sim e^{-k/\gamma}$ for some $\gamma>0$
which depends on the values of $p,q,r,s$.
See the Appendix for a derivation of this approximate distribution.

\paragraph{Overall degree distribution.}
The approximate analytic estimation of the expected degree distribution in the growth process 
and the fact that the MST used initially has a maximum degree of five
suggests that, for large enough $N$, the actual overall distribution of the node degree $K$ 
also has an approximately exponentially decaying tail:
\begin{align}
	P(K=k) &\sim e^{-k/\gamma}\quad\text{for large~}k.
\end{align}
Indeed, our simulations below show a good fit of the actual large-degree distributions with this form.

For small $k$, the distribution may already be monotonically decreasing,
but may also peak at $k=2,3,$ or $4$ for several reasons depending on $N_0,p,q$ and $s$:
If $N_0$ is large, the peak in the MST distribution may cause a peak in the overall distribution
(Fig.\ \ref{fig:polars},a,c,f). Further, if $s$ is large enough, this will also add to the share of degree-two nodes.
Eqns.\ \ref{eqn:p1},\ref{eqn:p2} from the Appendix imply that $p_2 > p_1$ iff 
\begin{align}
	\frac{s}{1-s} + s(1+p+2q) + 3p + 2pq + p^2 > 1.
\end{align}
Sufficient conditions for this are $s>(3-\sqrt{5})/2\approx 0.4$ or $p>(\sqrt{13}-3)/2\approx 0.3$.

\paragraph{Estimation of the exponential decay parameter from data.}
Assuming an exponentially decaying tail (i.e., a shifted geometric distribution)
for $k>k_0$ with some $k_0$,
one can easily show (by differentiating the log-likelihood function) 
that the maximum likelihood (ML) estimates of $\gamma$ 
given some observed degrees $k_1,\dots,k_N$ are given by
\begin{align}
	\hat\gamma  &= - 1/\log\left(1 - \frac{1}{\frac{\sum_{i,k_i>k_0}k_i}{|\{i:k_i>k_0\}|} - k_0}\right),
\end{align}
that is, by the relative frequencies of low degrees and by a function of the mean of the large degrees
(similar to the ML estimation of a geometric distribution, see \cite{Bracquemond2002}).
Alternatively to the ML estimator $\hat\gamma$,
one can also use the popular estimator $\hat\gamma'=-1/s$
where $s$ is the slope of a linear fit of the logarithmic decumulative degree distribution function of $k>k_0$.
Note that $\hat\gamma'$ places more weight on the rare large degrees,
while $\hat\gamma$ depends more on the bulk of the distribution,
which explains why typically $\hat\gamma>\hat\gamma'$ 
whenever the initial part of the distribution function is concave as in Fig.\ \ref{fig:polars}.

\subsection{Examples}

\paragraph{Example similar to Western US  power grid.}

\begin{figure}[htp]
\centering
\subfigure[]{\includegraphics[width=.45\textwidth]{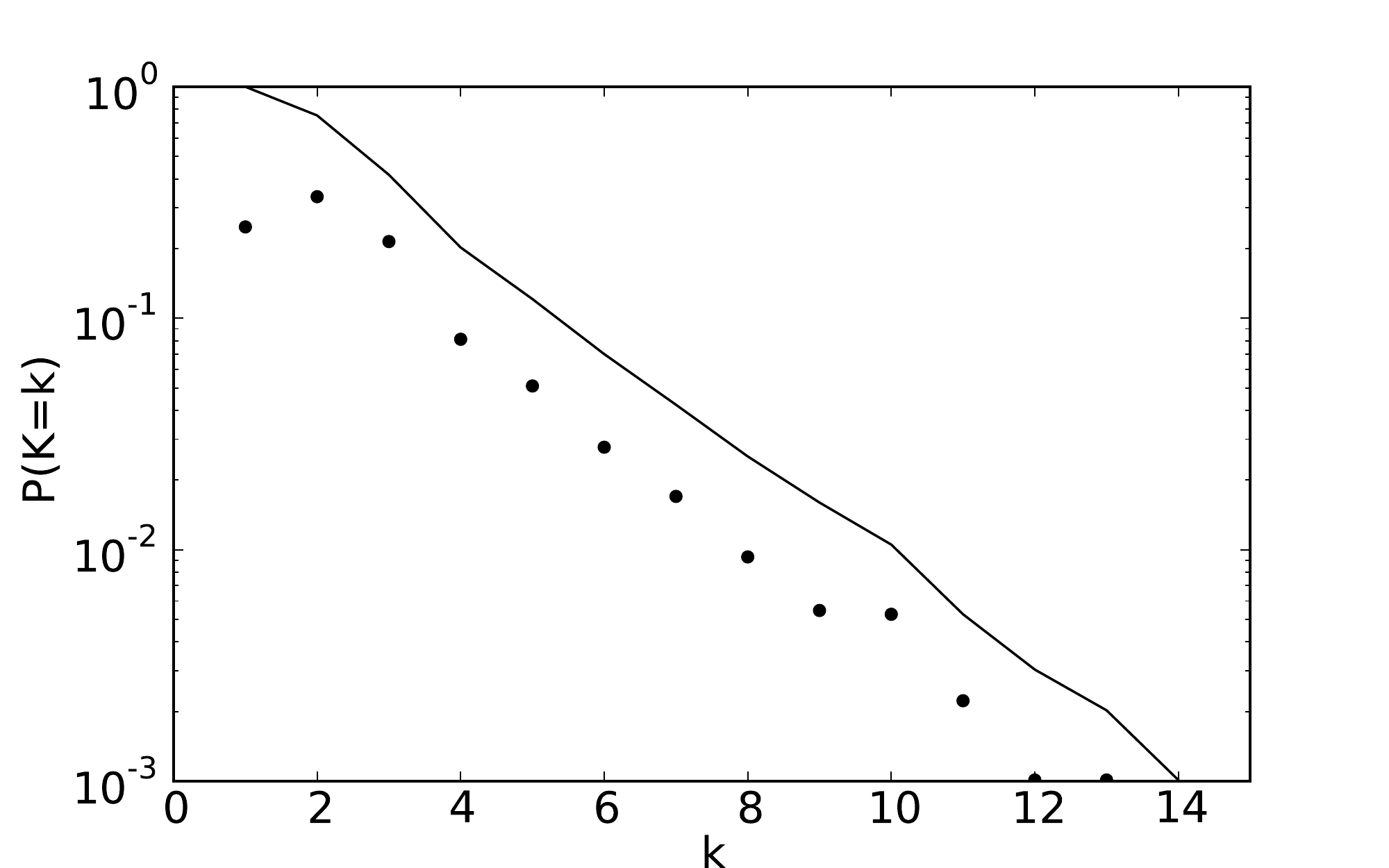}\label{fig:power}}
\subfigure[]{\includegraphics[width=.45\textwidth]{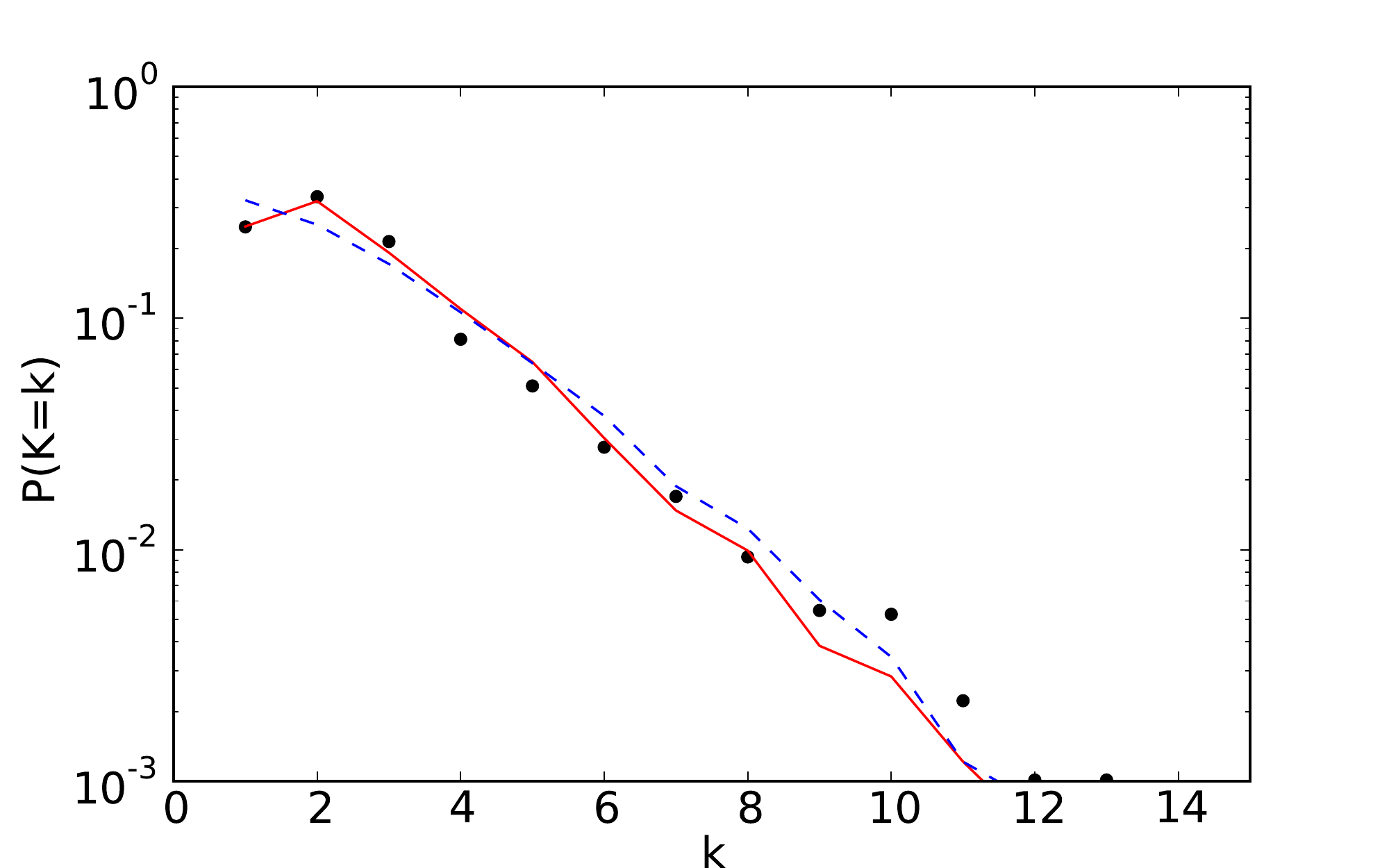}\label{fig:model}}
\caption{(a) probability mass function (dotted) and decumulative distribution (solid) of degrees from the Western US data set \cite{Watts1998} (b) model with ${N_0,p,q,r,s}={100, 0.01, 0.32, 1, 0}$ (dashed line) and ${N_0,p,q,r,s}={50,0.03,0.44,0.3,0.28}$ (solid line) and Western US data (dotted).}
\end{figure}

\begin{table}
\caption{Comparison of the Western US power grid data set (top row) and two realizations of our model with $(N_0,p,q,r,s)=(100,0.01,0.32,1,0)$ (middle row) and
$(N_0,p,q,r,s)=(50,0.03,0.44,0.3,0.28)$ (bottom row). ``cc'' denotes the global clustering coefficient and ``aspl'' the average shortest path length. All three networks have $N=4941$ nodes.}
\centering
\label{tab:wus}       
\begin{tabular}{llllll}
\hline\noalign{\smallskip}
M & $\bar k$  & $\hat\gamma'$ & $\lambda_2$ & cc & aspl\\
\noalign{\smallskip}\hline\noalign{\smallskip}
6594 & 2.67 & 2.12 & 0.001 & 0.08 & 18.99\\
6597 & 2.67 & 1.50 & 0.003 & 0.09 & 12.58\\
6622 & 2.68 & 1.33 & 0.001 & 0.12 & 16.54\\
\noalign{\smallskip}\hline
\end{tabular}
\end{table}

To illustrate realizations of our model, we have chosen the popular data set of the Western US  power grid \cite{Watts1998}, see Fig.\ \ref{fig:power}.
With the parameters $(N_0,p,q,r,s)=(100,0.01,0.32,1,0)$ in Fig.\ \ref{fig:model} we are able to fit the observed mean degree and the slope of the exponential decay very well (see Tab.\ \ref{tab:wus}). Note that this is a realization in the absence of splitting ($s=0$), what might serve as an explanation for  the absence of a maximum  at degree two, also this leads to a smaller average shortest path length as will be discussed below in more detail. For comparison we also show a different 
realization with $(N_0,p,q,r,s)=(50,0.03,0.44,0.3,0.28)$ including the splitting of lines.

\paragraph{Polar examples.}

\begin{figure}\begin{center}\begin{tabular}{ccc}%
(a) &(b) &(ENTSO-E region 1) \\
\includegraphics[width=0.25\textwidth]{images/polar_a400}
& \includegraphics[width=0.25\textwidth]{images/polar_b400} 
& \includegraphics[width=0.25\textwidth,height=0.25\textwidth]{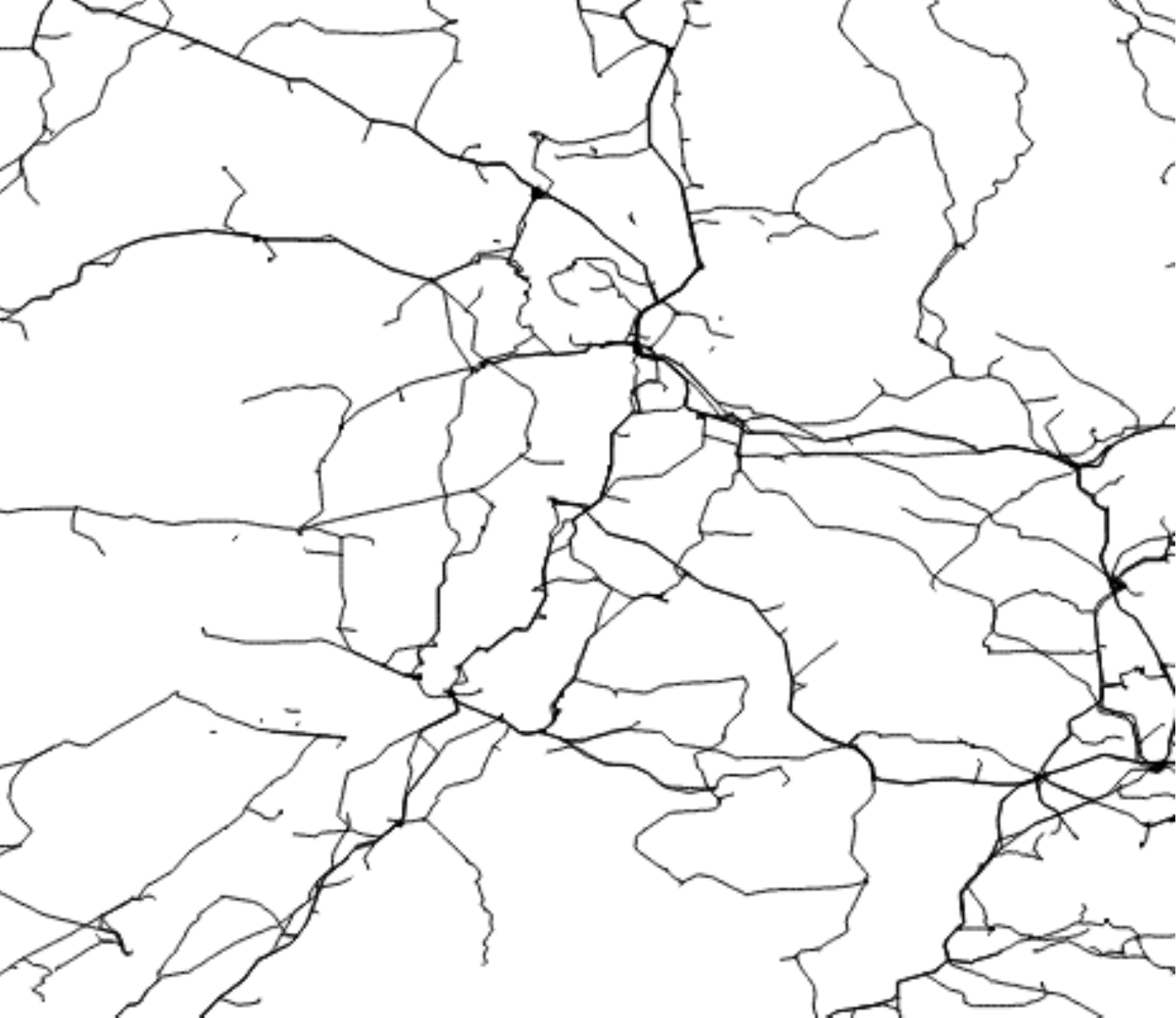}\\

\includegraphics[width=0.3\textwidth,height=0.2\textwidth]{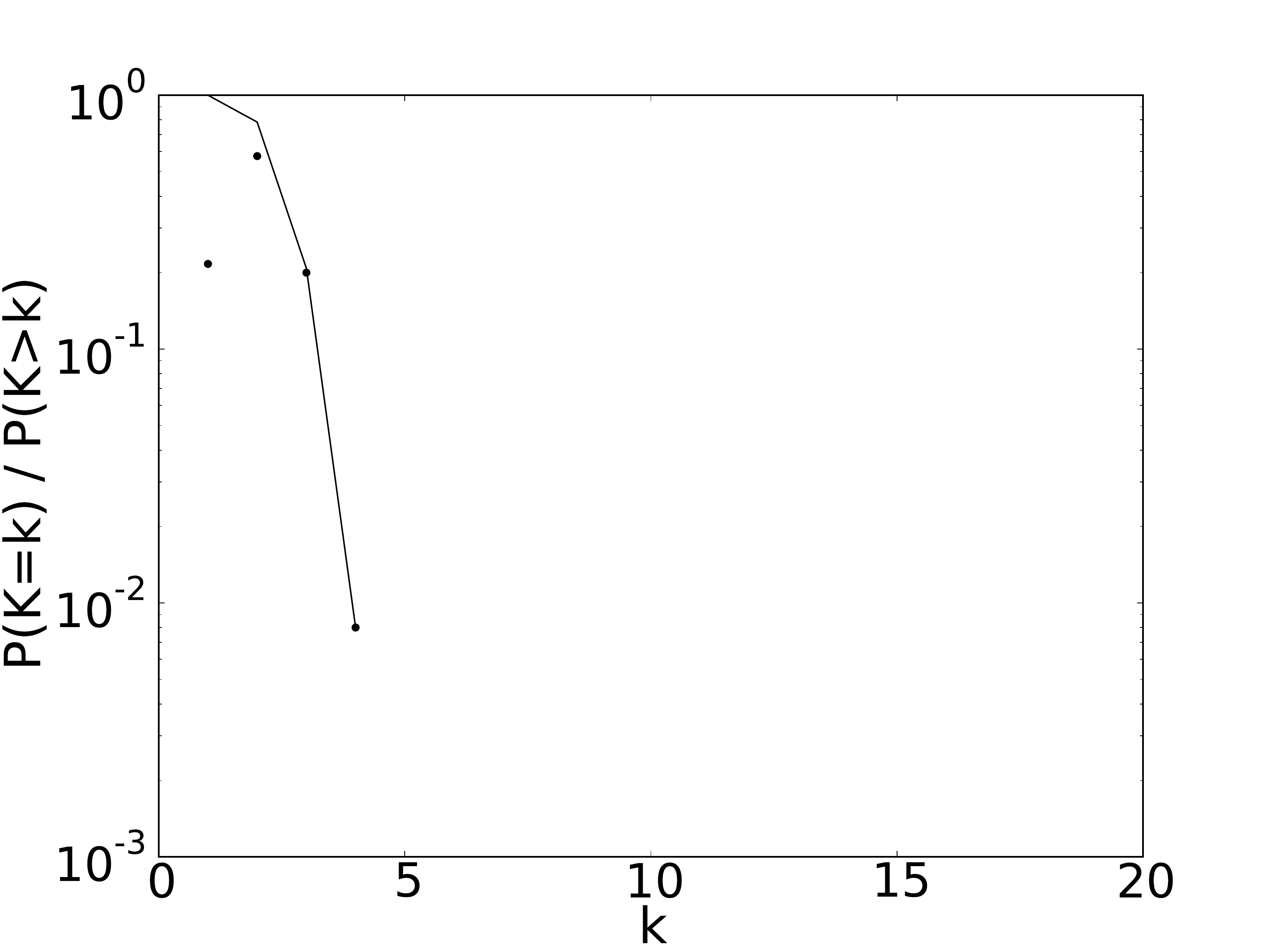}
&\includegraphics[width=0.3\textwidth,height=0.2\textwidth]{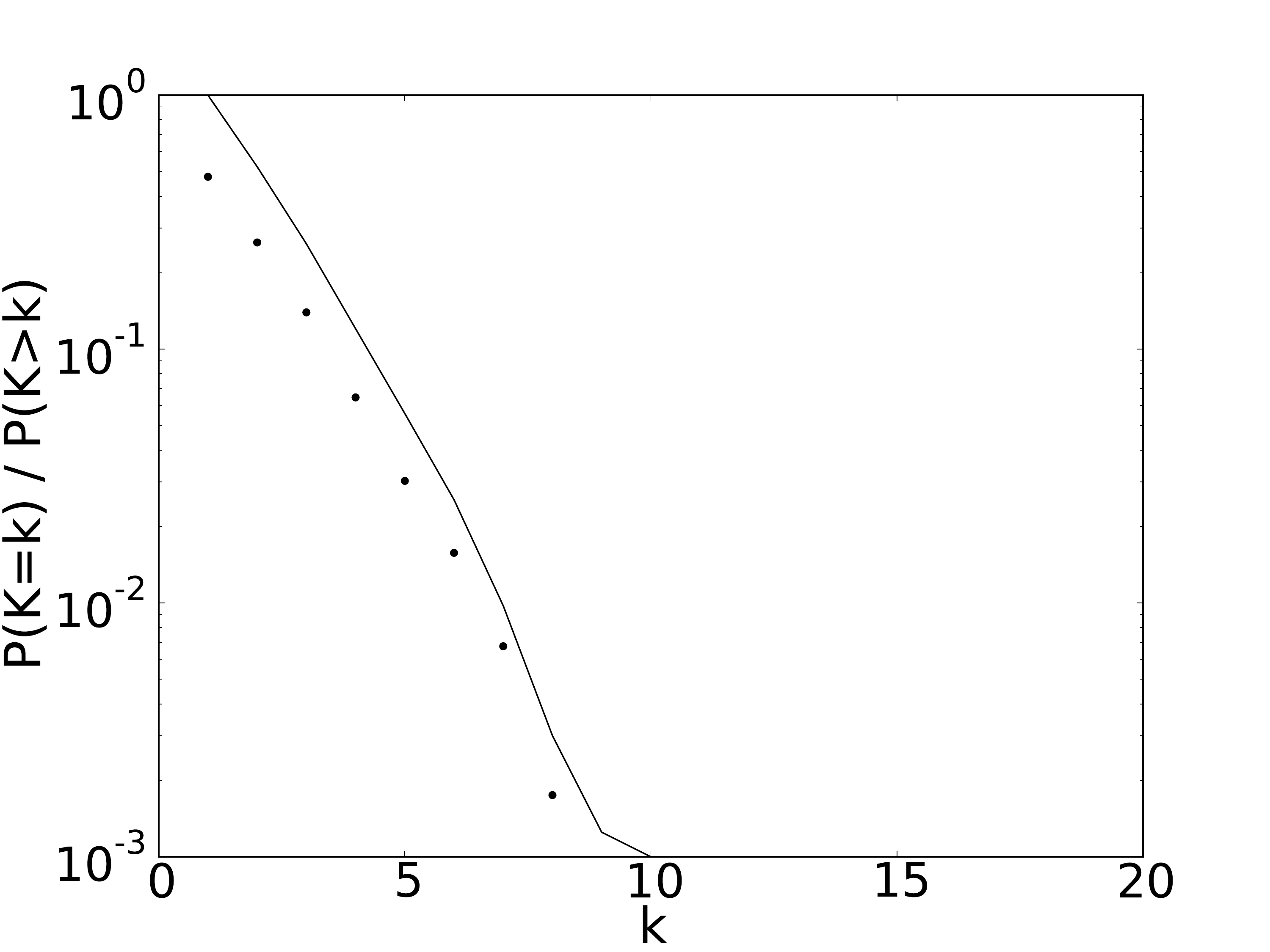}
&\includegraphics[width=0.3\textwidth,height=0.2\textwidth ]{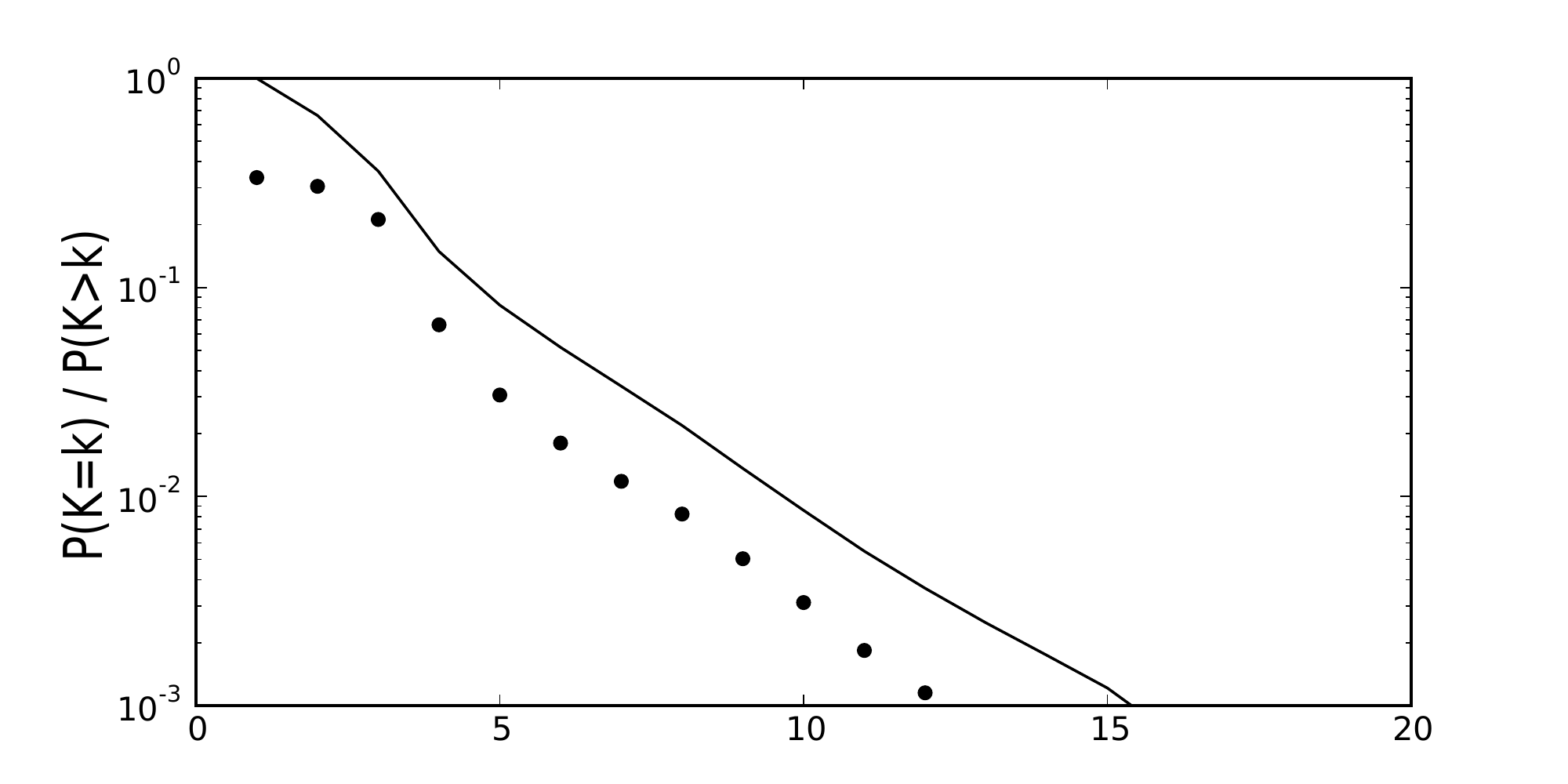}\\

(c) & (d) & (e)\\ 
\includegraphics[width=0.25\textwidth]{images/polar_c400}
&\includegraphics[width=0.25\textwidth]{images/polar_d400}
&\includegraphics[width=0.25\textwidth]{images/polar_e400}\\

\includegraphics[width=0.3\textwidth,height=0.2\textwidth]{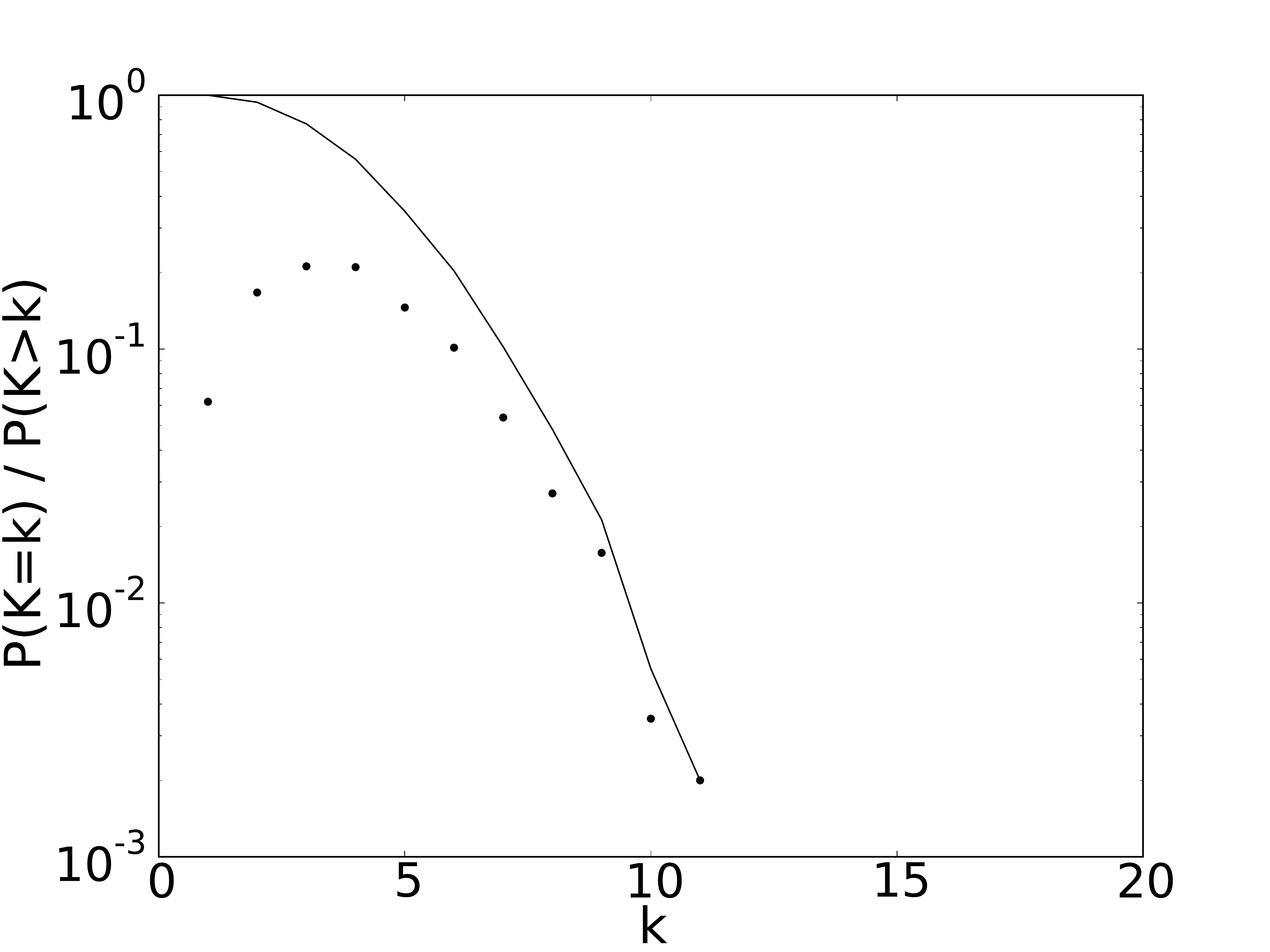}
&\includegraphics[width=0.3\textwidth,height=0.2\textwidth]{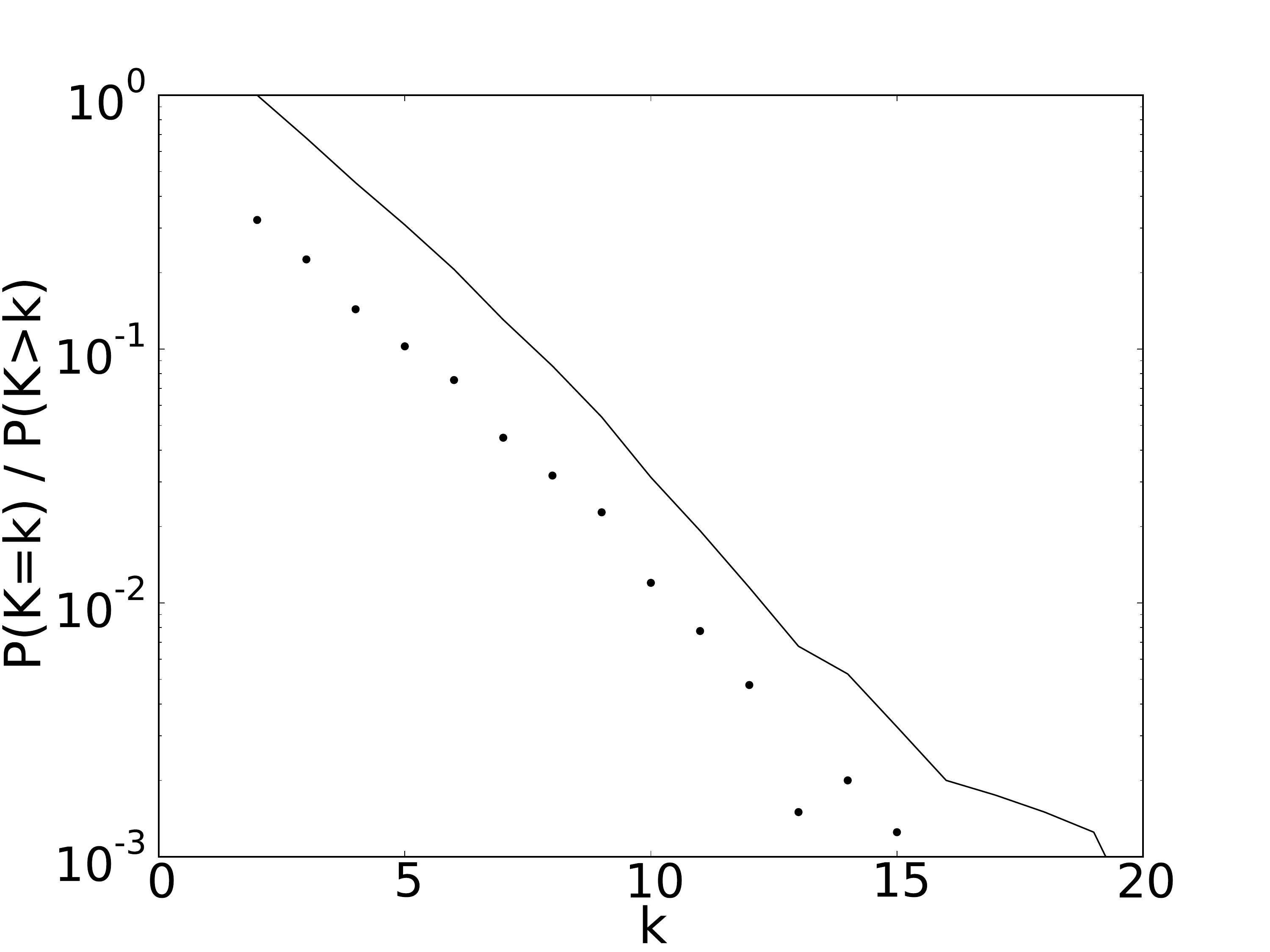}
&\includegraphics[width=0.3\textwidth,height=0.2\textwidth]{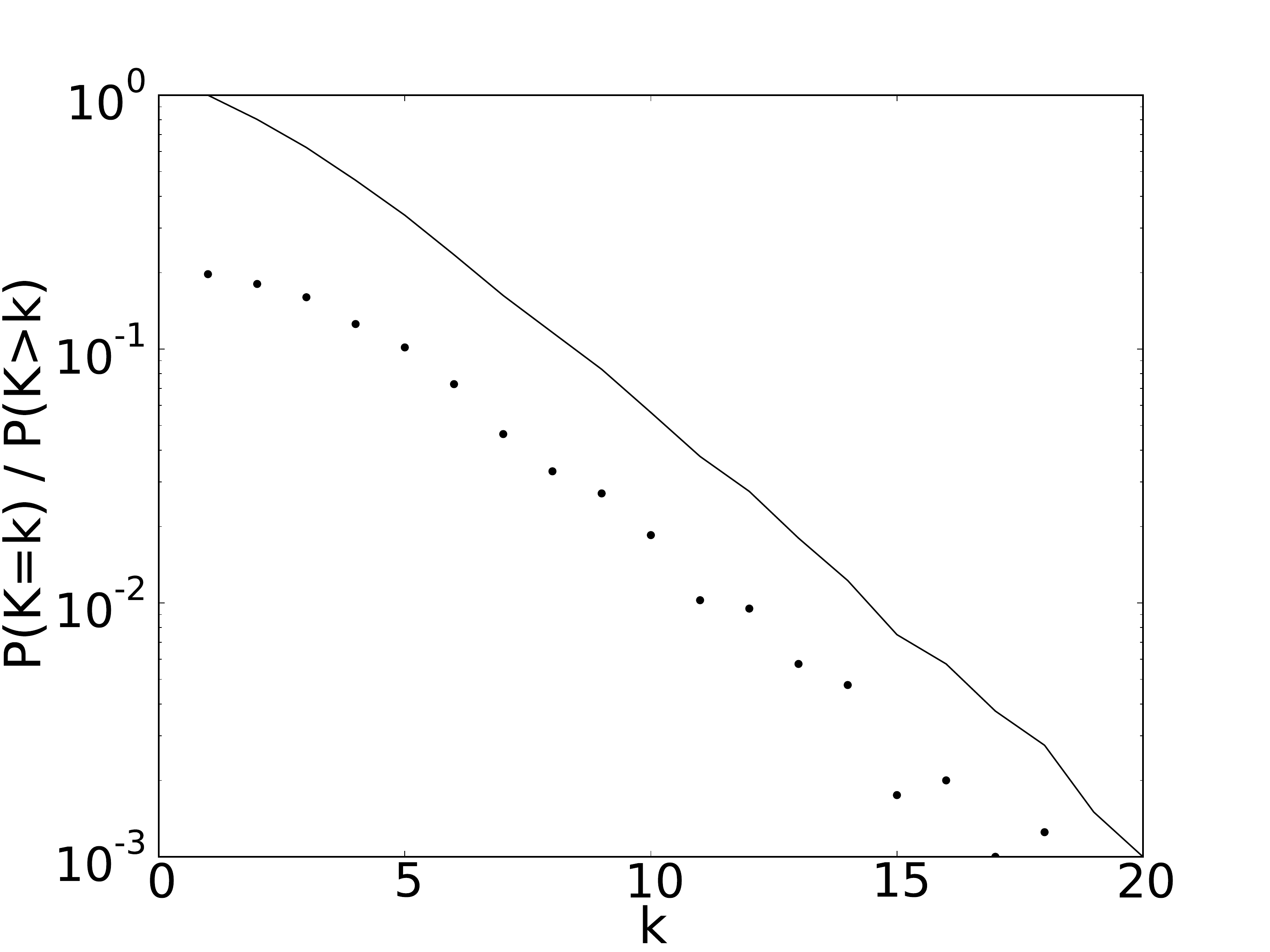}\\

(f)&(g)&(h)\\
\includegraphics[width=0.25\textwidth]{images/polar_f400}
&\includegraphics[width=0.25\textwidth]{images/polar_g400}
&\includegraphics[width=0.25\textwidth]{images/polar_h400}\\

\includegraphics[width=0.3\textwidth,height=0.2\textwidth]{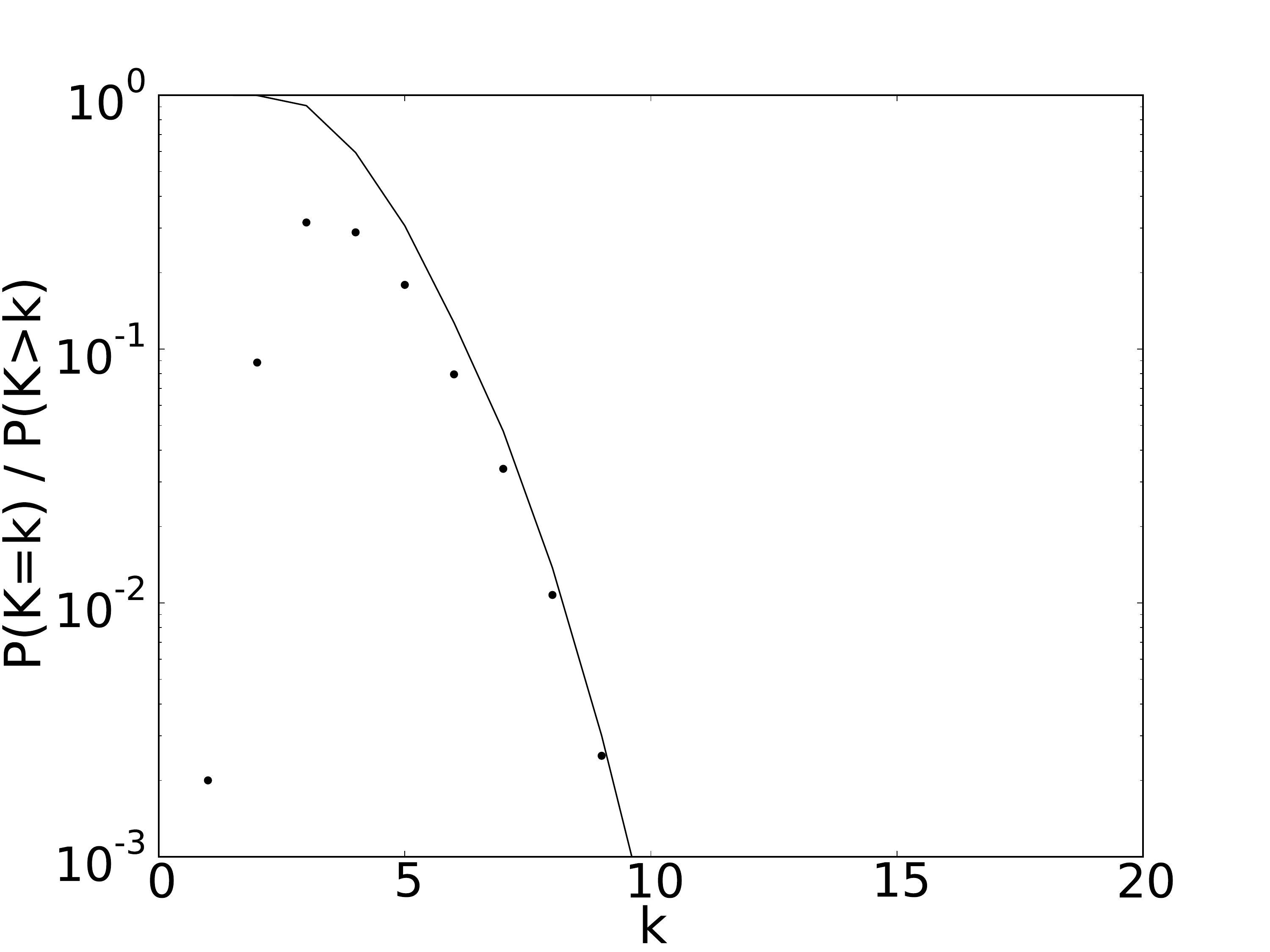}
&\includegraphics[width=0.3\textwidth,height=0.2\textwidth]{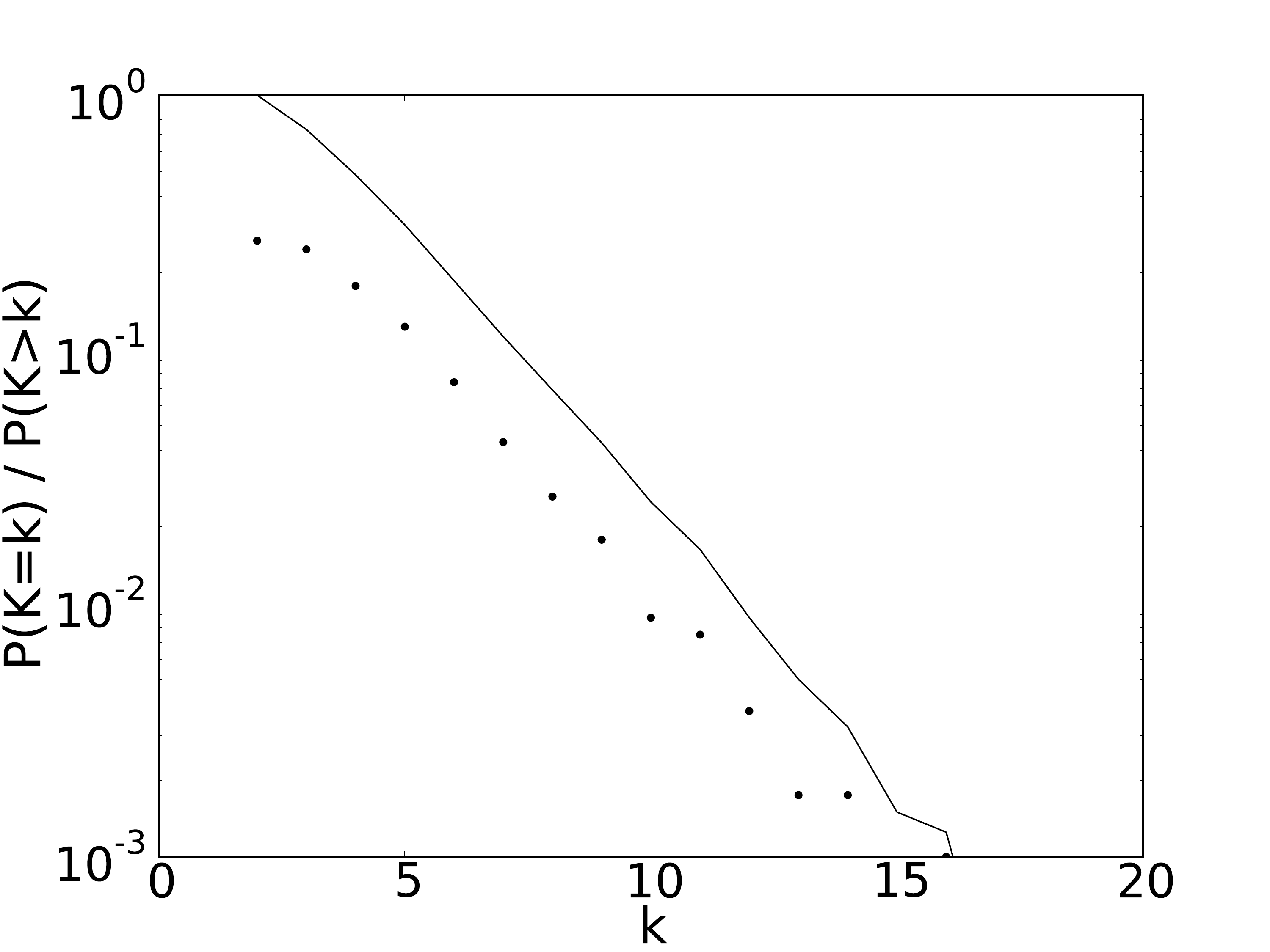}
&\includegraphics[width=0.3\textwidth,height=0.2\textwidth]{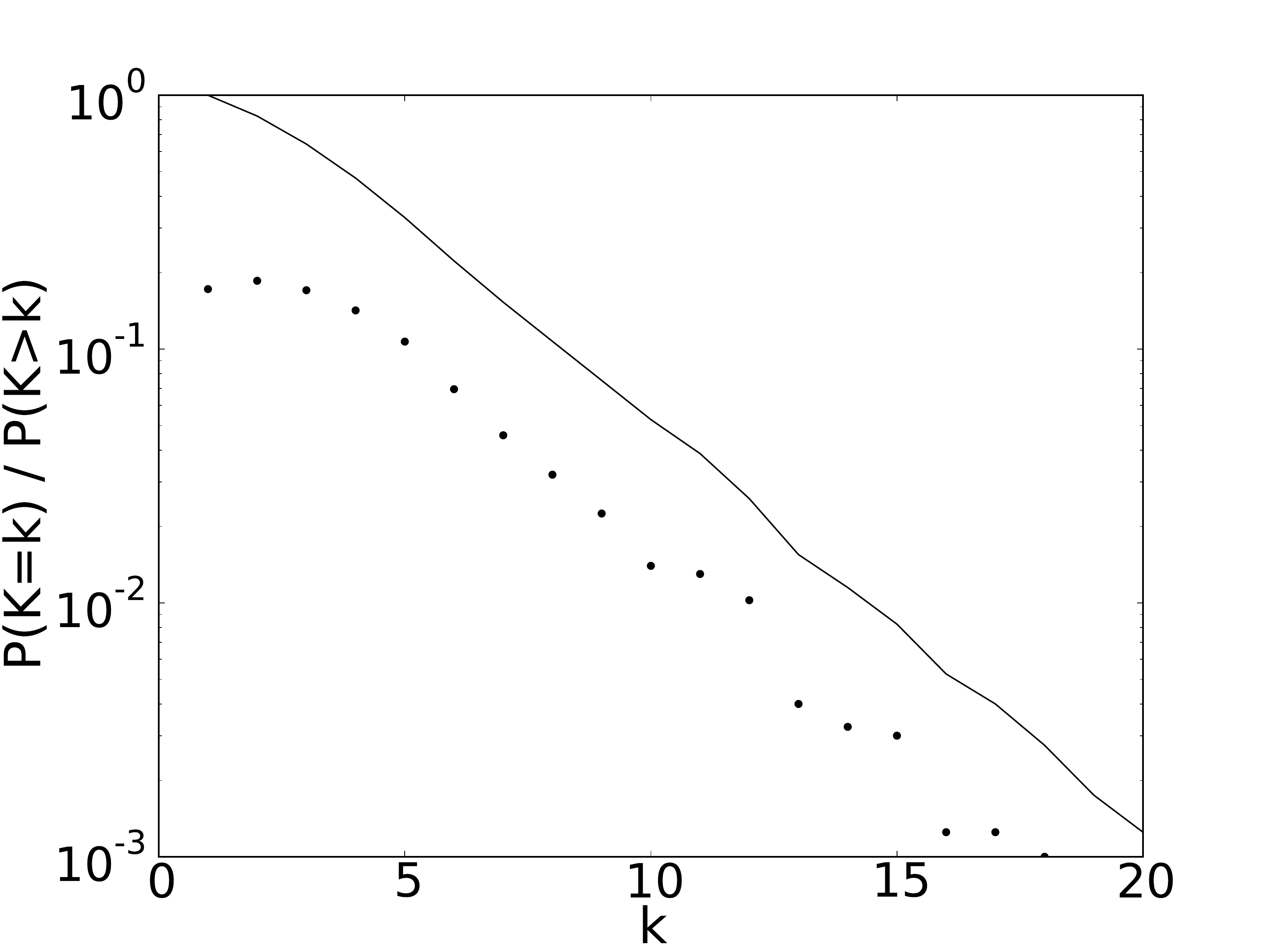}\\
\end{tabular}\\
\end{center}
\caption{(Color online) 
Polar examples of random grid topologies with $N=400$ nodes and degree distributions for $N=4000$ nodes,
generated with our model ($s=0$ in all plots).
Relative frequencies $P(K=k)$ (dots) and
decumulative distribution function $P(K>k)$ (solid lines).
Links added in step I2, I3, G2, G3, or G4 are plotted in black, brown, blue, red, or orange, respectively.
Left column: whole grid constructed with minimum spanning tree in initial phase ($N_0=N$); 
Middle and right columns: no initial phase, whole grid successively grown 
($N_0=1$, middle: $p\ge 0$, $q=0$; right: $p=0$, $q\ge 0$). 
Top row: tree-shaped grids with mean degree $\approx 2$ ($p=q=0$);
Middle row: grids with redundant links minimizing spatial distance ($p+q=1$, $r=0$);
Bottom row: grids with redundant links maximizing a redundancy/cost trade-off ($r=1$).
Top right: Detail of real-world power grid in ENTSO-E region 1 (mainland Europe) for comparison.}
\label{fig:polars}
\end{figure}

Fig.\ \ref{fig:polars} shows examples of grids and degree distributions 
generated with our model for ``polar'' choices of $p,q,r\in\{0,1\}$
with $s=0$.
The plotted topologies picture networks with $N=400$ nodes, 
and each degree distribution shows the result for one individual network with a larger number of $N=4000$ nodes
so that the tail shape can be seen more easily. An exponential decay appears linearly in the semi-logarithmic plot.


In (a), $N_0=N$ and $p=q=r=0$, hence the grid is simply the MST of $N$ random locations,
with no redundant or long-range links, a mean degree of two and a flat degree distribution peaking at degree two.
In (b), $N_0=1$ instead, hence the grid is a tree grown by adding one node and a link to the closest existing node one at a time,
resulting in some long-range links and 
an approximately exponentially decaying degree distribution with $\gamma$ estimated as $\hat\gamma=1.24$ or $\hat\gamma'=1.12$.
In (c) and (f), $N_0=N$ again, resulting in a MST as the skeleton grid,
but with $p+q=1$, resulting in redundant links
either minimizing spatial distance ((c), $r=0$), 
leading to a clustering of links in regions which received more nodes by chance; 
or maximizing the quotient between network and spatial distance ((f), $r=1$),
leading to a mesh-like structure with almost no remaining degree-one nodes
and an exponential decay with $\hat\gamma=1.02$ or $\hat\gamma'=0.78$.
In (d), $N_0=1$, $p=1$, $q=0$, and $r=0$,
so that the grid is grown by connecting each new node with the two closest existing nodes,
resulting in an exponential decay with $\hat\gamma=2.30$ or $\hat\gamma'=2.10$.
In (e), $p=0$ and $q=1$ instead,
hence in each growth step both the new node and some existing node 
are linked to their respective closest other existing node,
resulting in an exponential decay with $\gamma$ estimated as $\hat\gamma=2.86$ or $\hat\gamma'=2.67$.
In (g), $N_0=1$, $p=1$, $q=0$, and $r=1$, so that in contrast to figure (d) the structure appears to be more mesh-like with more redundancy. 
Again this is because a new node is always linked to two more, giving equal weight to spatial and network distance.
Lastly in (h), $N_0=1$, $p=0$, $q=1$, and $r=1$, so that in every step a new node will connected to exactly one existing node,
 with $q=1$ meaning that also in every step two further nodes become directly linked. While this scenario still produces more 
 long-ranged connections as in (e), the structure becomes locally more meshed.
Additionally we picture a snapshot of the ENTSO-E region 1 network of continental Europe for comparison. ENTSO-E stands for ``European 
Network of Transmission System Operators for Electricity'', the successor of the Union for the Co-ordination of Transmission of Electricity (UCTE).
ENTSO-E is intended to ensure the coordination of network operation among the members, make agreements on network codes and to improve the integration between EU member state's markets.

\subsection{Monte-Carlo simulation and numerical test of the degree distribution}\label{subsec:monte-carlo}
\begin{figure}[htp]
\centering
\subfigure[]{\includegraphics[width=.33\textwidth]{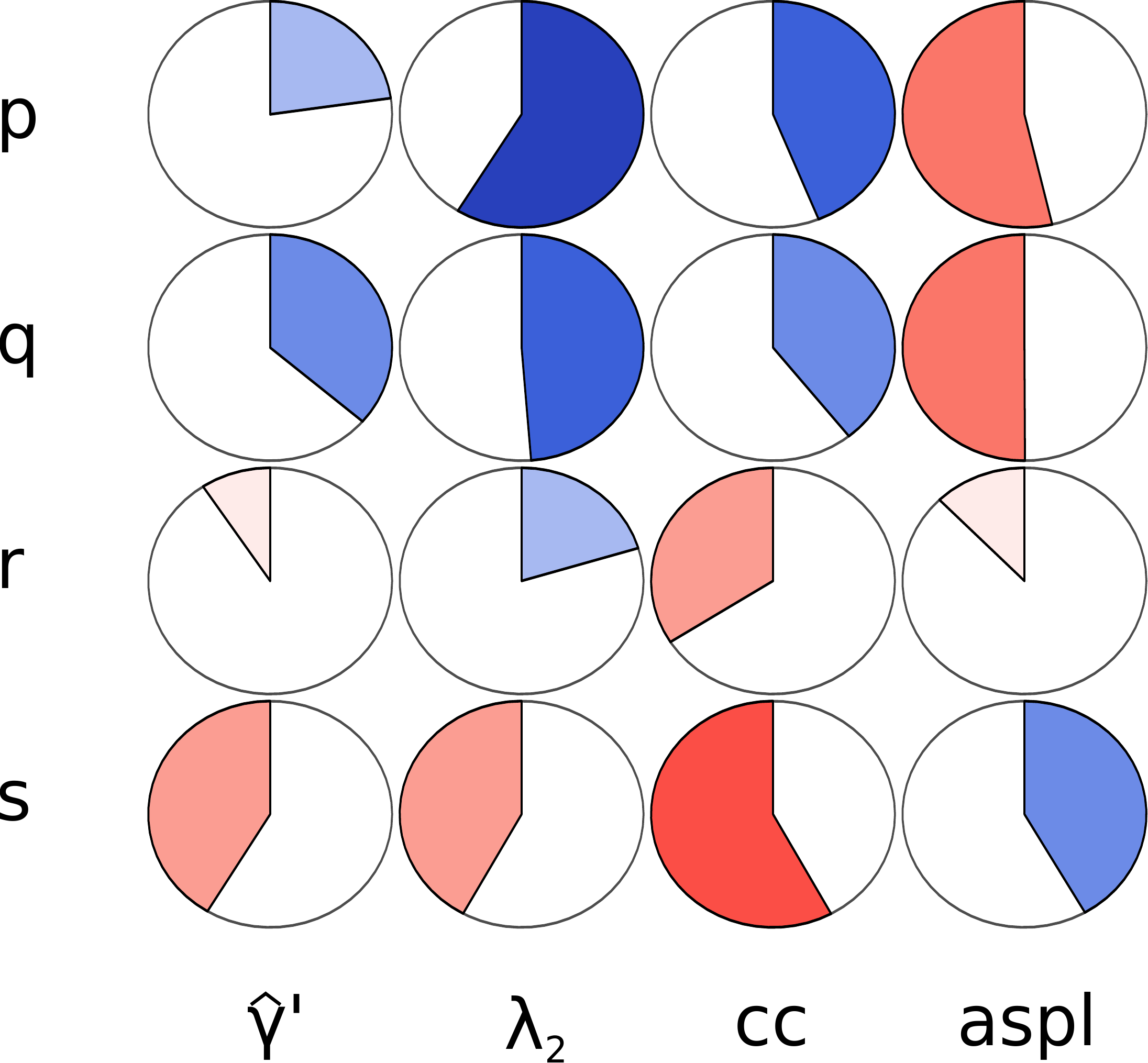}\label{fig:corr}}
\subfigure[]{\includegraphics[width=.45\textwidth]{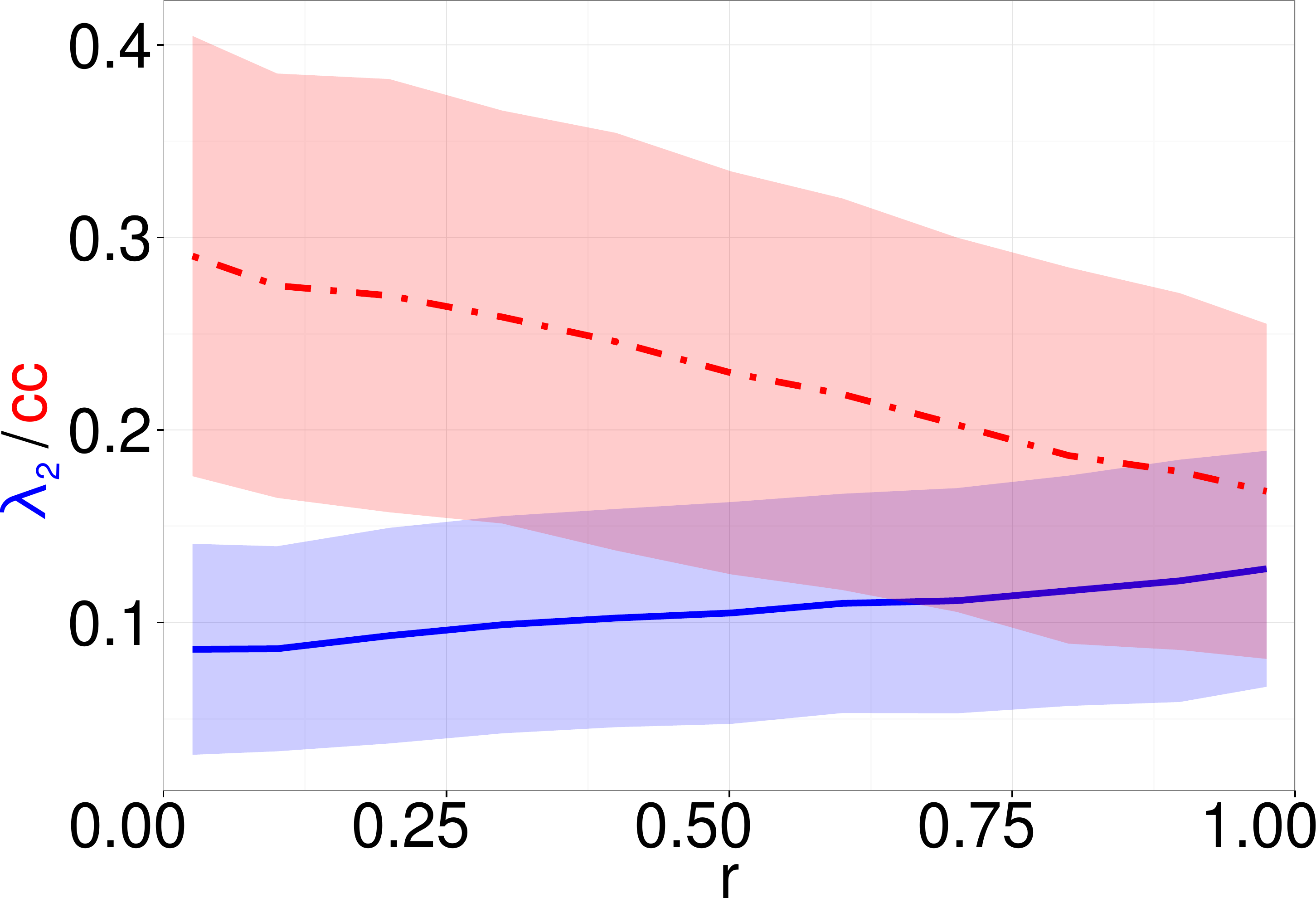}\label{fig:redundancy}}
\subfigure[]{\includegraphics[width=.45\textwidth]{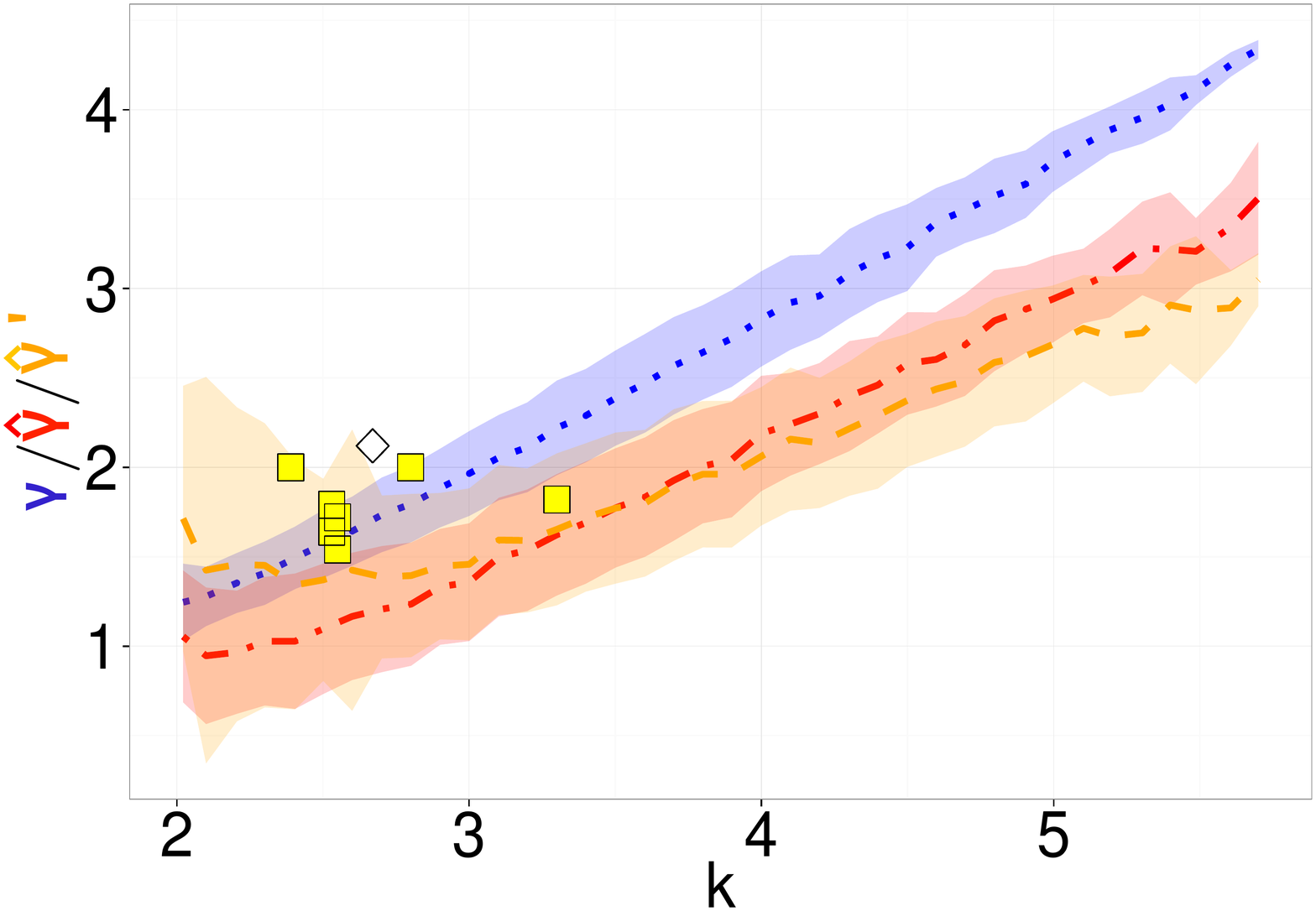}\label{fig:gammas_vs_k}}
\subfigure[]{\includegraphics[width=.45\textwidth]{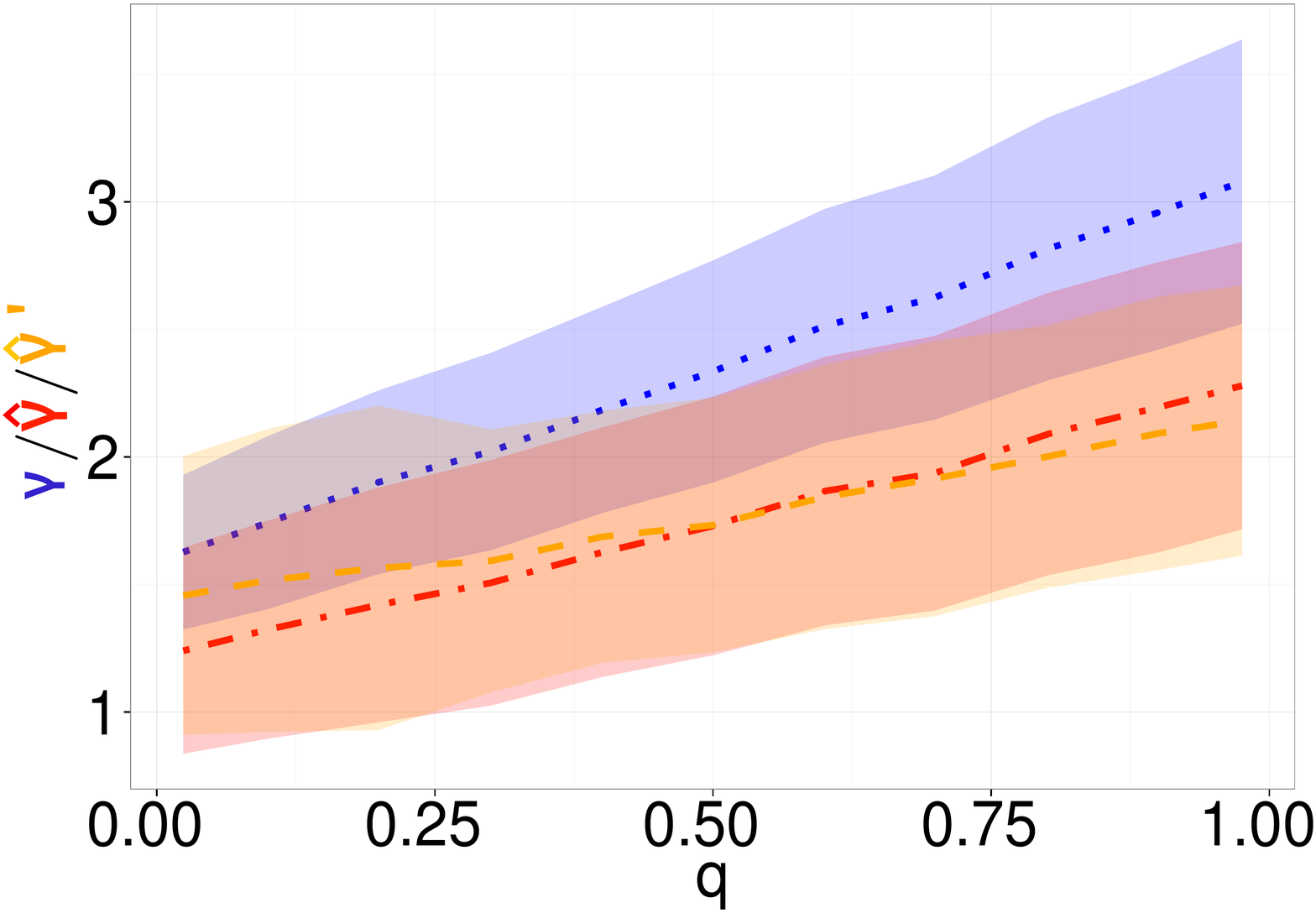}\label{fig:gammas_vs_q}}
\caption{(Color online) \textbf{(a)} Correlation matrix of the model parameter $p,q,r,s$ vs the slope $\hat\gamma'$ of the exponential decay, $\lambda_2$, the global clustering coefficient ``cc'' and the average shortest path length ``aspl''. Circles are filled proportional to the magnitude of the correlation, positive values are pictured in blue, negative values in red. Pictured are further: \textbf{(b)} averaged values of $\lambda_2$ (solid) and the global clustering coefficient (dashed) given the trade-off parameter $r$, \textbf{(c)} averaged values of the analytic estimate of $\gamma$ (dotted), $\hat{\gamma}$ (dot-dashed) and $\hat{\gamma}'$ (dashed) for realizations with different mean degree $k$ and \textbf{(d)} same as in (c) for realizations with different $q$. Shaded areas indicate one standard deviation. The diamond marks the Western US power grid, the yellow squares mark data from \cite{Pagani2013}.}
\label{fig:constN}
\end{figure}

\begin{figure}[htp]
\centering
\subfigure[]{\includegraphics[width=.45\textwidth]{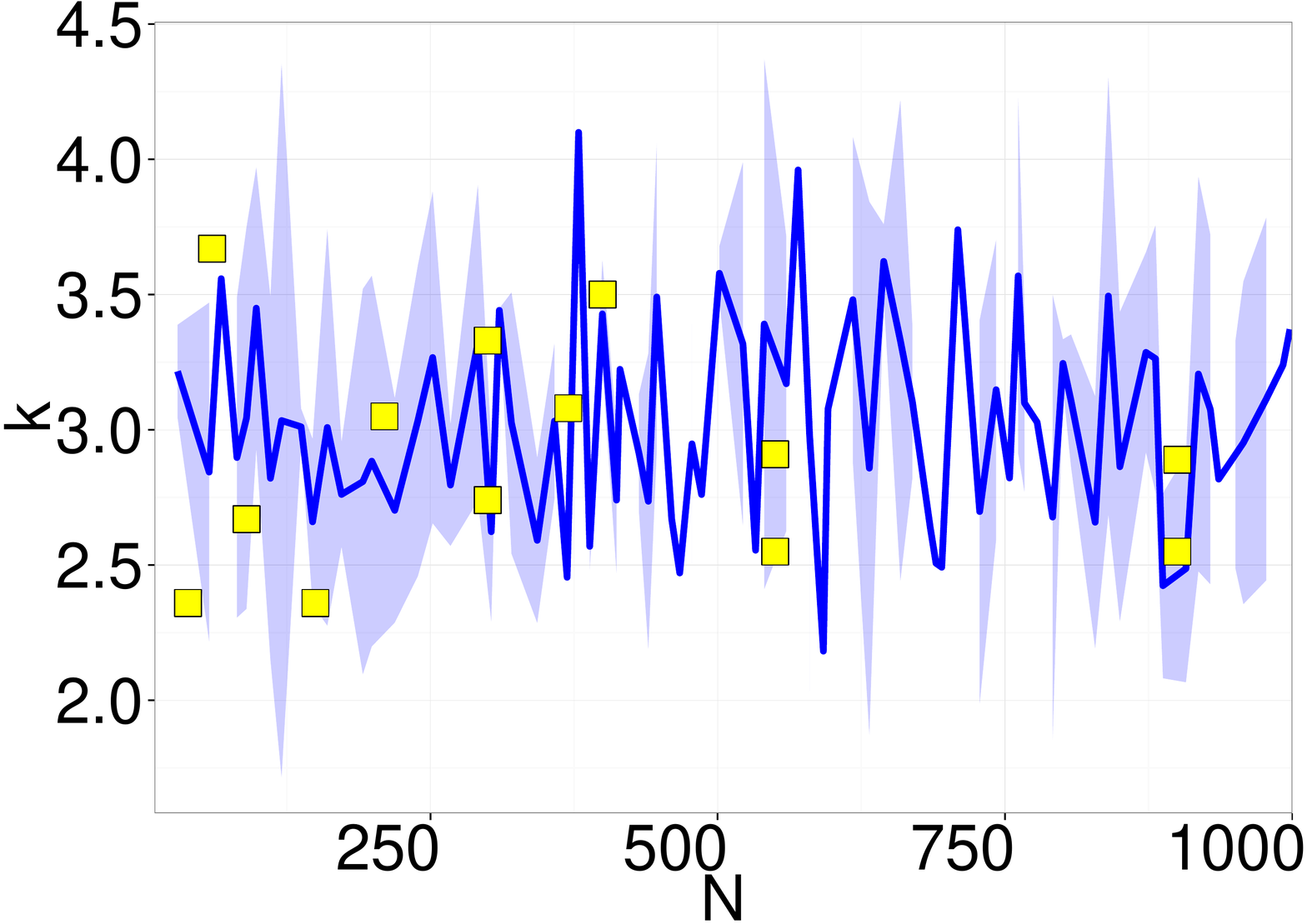}\label{fig:k_vs_N_DATA_p_or_q_less_0p1}}
\subfigure[]{\includegraphics[width=.45\textwidth]{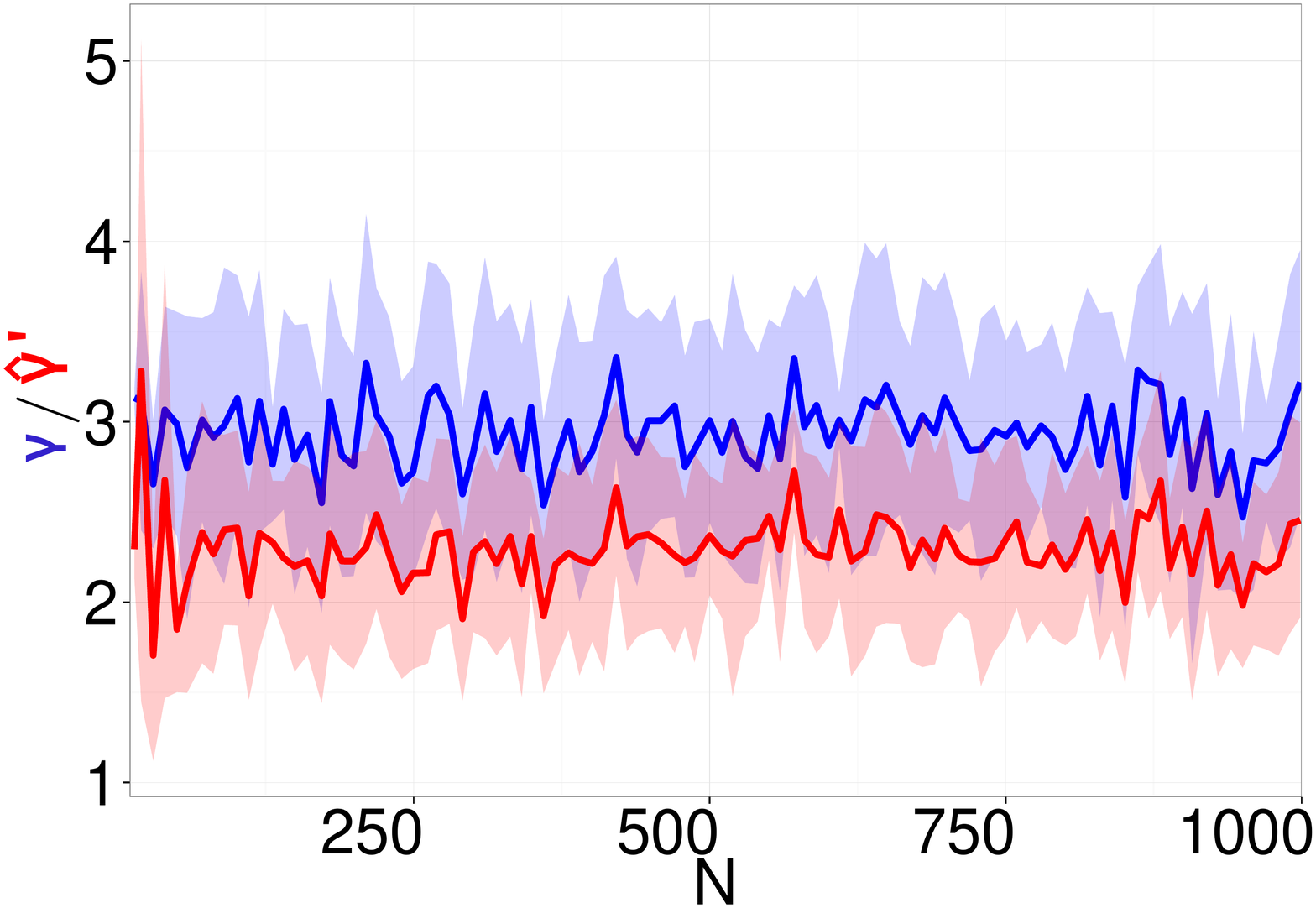}\label{fig:gamma_and_slope_vs_N}}
\subfigure[]{\includegraphics[width=.45\textwidth]{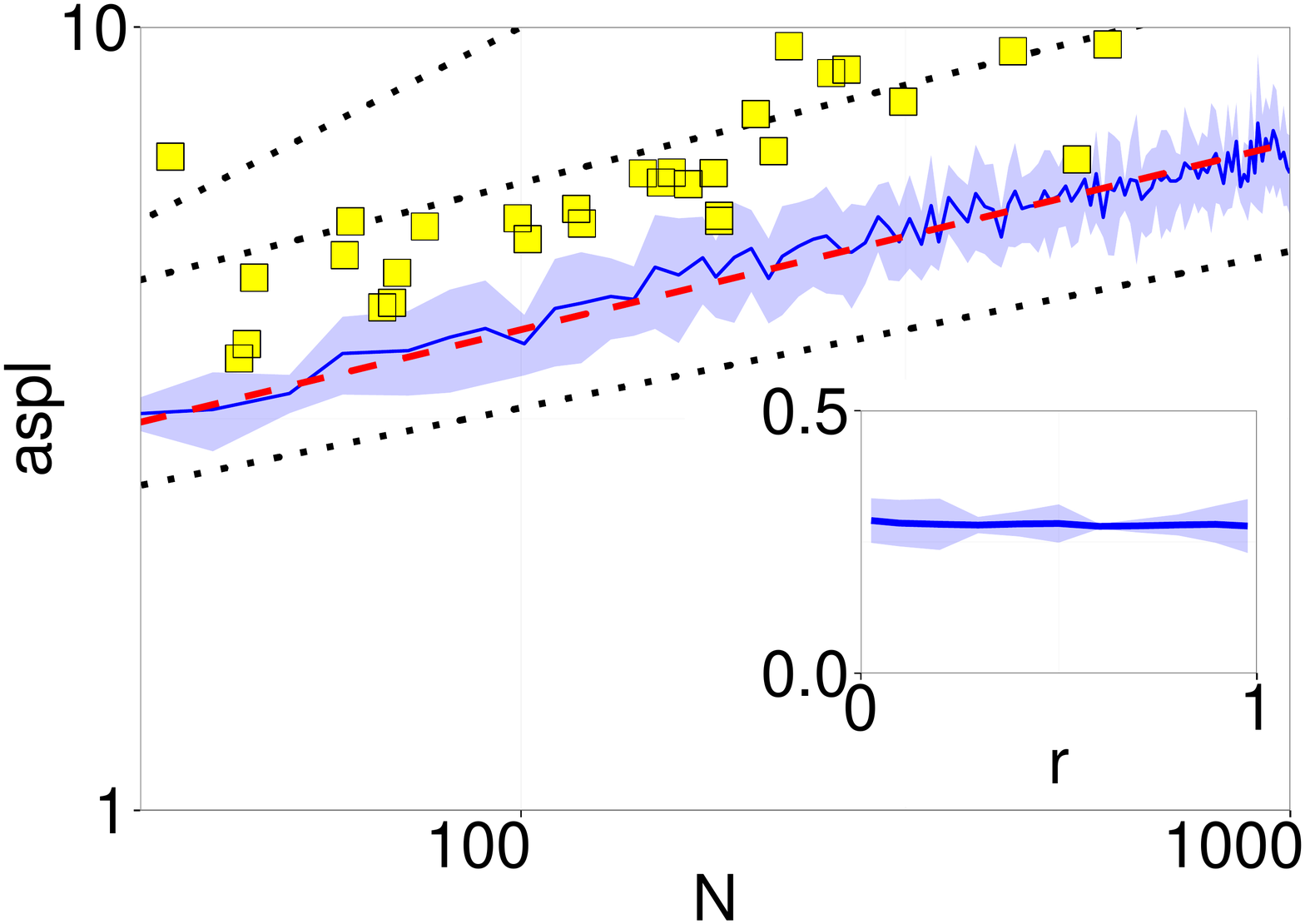}\label{fig:aspl_vs_N_INSET_DATA}}
\subfigure[]{\includegraphics[width=.45\textwidth]{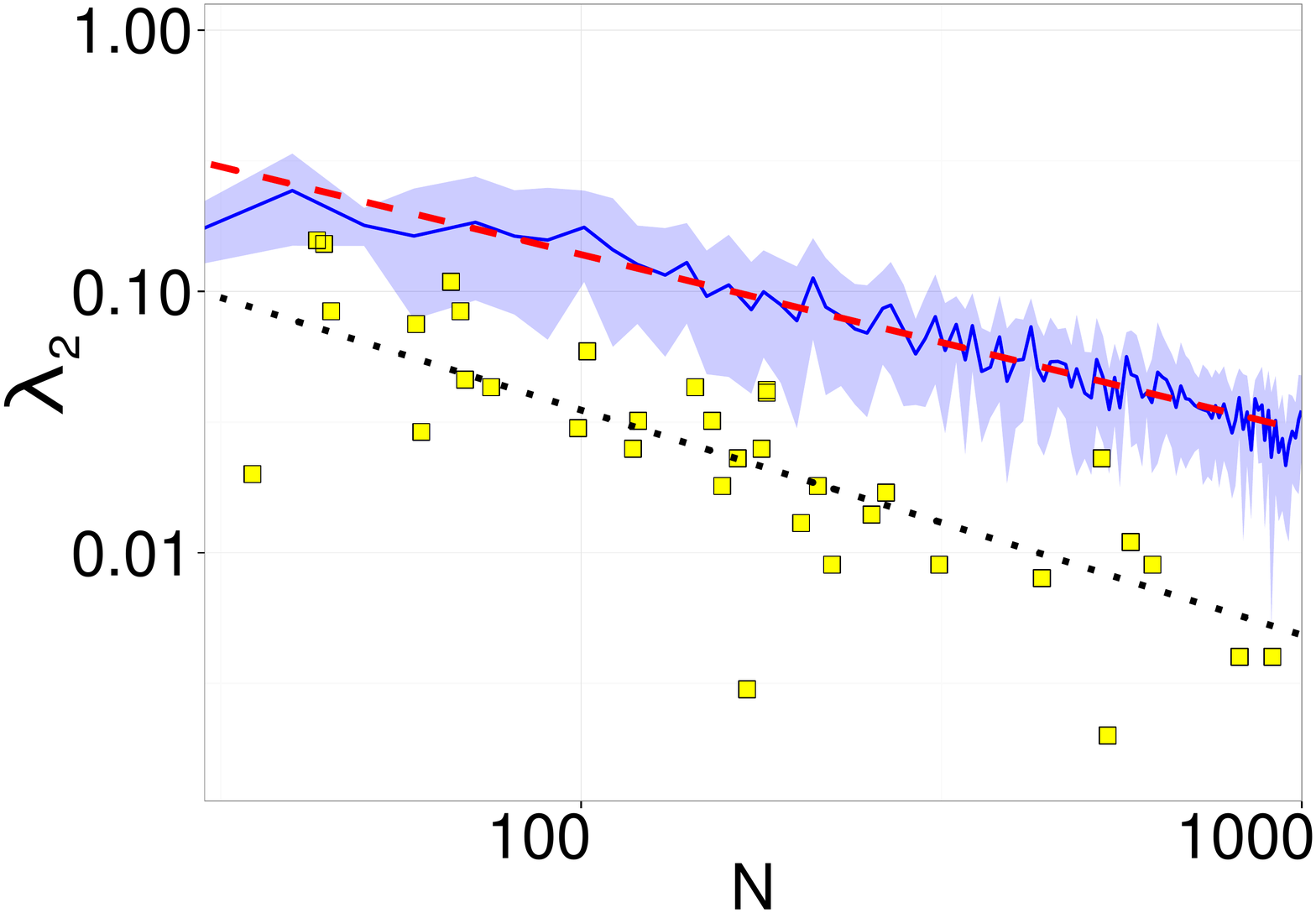}\label{fig:lambda2_vs_N}}
\caption{(Color online) Pictured are: 
\textbf{(a)} the mean degree $k$ vs. $N$ for a subsample with $p$ \emph{or} $q$ less than $0.1$ (yellow squares mark data from \cite{Pagani2013});
\textbf{(b)} $\gamma$ (top) and $\hat\gamma'$ (bottom) vs $N$ and 
\textbf{(c)} a double-logarithmic plot of the average shortest path length ``aspl'' (solid) vs. $N$. 
The dotted lines have a slope of $1/2$,$1/4$ and $1/5$ from top to bottom, 
the yellow squares mark data for the ENTSO-E member states (see Appendix). 
The inset shows $(\log\, aspl / \log\, N)$ vs. $r$. \textbf{(d)} $\lambda_2$ vs. $N$ (solid), 
where the dashed line indicates a linear regression to the model data and the dotted line a linear regression to the data of the ENTSO-E member states. Shaded areas indicate one standard deviation.}
\label{fig:varN}
\end{figure}

We performed two sets of Monte-Carlo simulations.
(i) an ensemble of 10000 model realizations with fixed node number $N=100$ ($N_0=1$) 
and randomly chosen $p,q,r\in[0;1]$ and $s\in[0;0.5]$ (see Fig.\ \ref{fig:constN})
and
(ii) an ensemble of 1000 realizations with $N\in [10;1000]$ ($N_0=1$) 
and randomly chosen $p,q,r\in[0;1]$, $s=0$ (see Fig.\ \ref{fig:varN}).

In Fig.\ \ref{fig:corr} we picture the correlation matrix of the model parameter $p,q,r,s$ vs. 
the estimated decay parameter $\hat\gamma'$, the algebraic connectivity $\lambda_2$, the global clustering coefficient 
and the average shortest path length aspl. 
It can be seen that the parameter $r$ mainly influences the algebraic connectivity and the clustering
since it parametrizes a cost-vs.-redundancy trade-off for the creation of new links. 
In Fig.\ \ref{fig:redundancy} it seems that there also happens a trade-off between global redundancy ($\lambda_2$) and local redundancy (clustering coefficient), 
where $\lambda_2$ is increasing with $r$ and vice versa.
Besides, we see from Fig.\ \ref{fig:corr}, that the average shortest path length is indeed increasing with $s$ in accordance to our above observation.

The influence of the parameters $p$ and $q$ is very similar (Fig.\ \ref{fig:corr}). 
From Figs.\ \ref{fig:gammas_vs_q} and \ref{fig:gammas_vs_k} we conclude that they for instance determine the mean degree and the slope of the exponential tail 
in an almost linear way on average. 
However it becomes obvious that there is a systematic deviation of our estimators $\hat\gamma$ and $\hat\gamma'$ 
from the analytically determined theoretical value of $\gamma$ (see the Appendix)
which remains to be explained theoretically so far.

Further on we studied the scaling behavior of some network measures (Fig.\ \ref{fig:varN}). 
Again we find an offset between the estimates $\hat\gamma'$ and $\gamma$ (Fig.\ \ref{fig:gamma_and_slope_vs_N})
but at least the observation supports our finding that the slope of the exponential decay does not depend on the network size $N$ (see Eqn.\ \ref{eqn:gamma}). 
Supporting the observation in Fig.\ \ref{fig:k_vs_N}, we find that the mean degree does not depend on $N$, if we choose appropriate parameter values, the range of model realizations matches the data very well (Fig.\ \ref{fig:k_vs_N_DATA_p_or_q_less_0p1}).

In Fig.\ \ref{fig:aspl_vs_N_INSET_DATA} we see that the scaling of the average shortest path length with the network size follows a $N^\delta$ dependence, 
where a linear regression (dashed line) yields $\delta\approx 0.24\pm 0.01$. 
As a guidance to the eye, we also pictured $\delta=0.5$ (dotted line) that would correspond to a network with only very local connections, 
whereas in these realizations of our model, the possibility of long-range connections leads to a weaker increase. 
For comparison we plotted data points for the ENTSO-E member states (see Appendix) which is well in the range of the model realization. 
Where the model yields to short path length, it might again be necessary to take the splitting of lines into account. 
From the plot of $(\log\, aspl / \log\, N)$ vs. $r$ we see that this scaling behavior does not depend on the degree of redundancy.
Contrary, the global clustering coefficient shows no dependence at all on $N$ in accordance to empirical findings (see Tab.\ \ref{tab:ucte})

In the last figure, we investigate the scaling of the algebraic connectivity with the network size (Fig.\ \ref{fig:lambda2_vs_N}). 
We find the $\lambda_2\sim N^{-\eps}$ where a linear regression yields $\eps\approx 0.73\pm0.05$ similar to the result from the ENTSO-E data set $\eps\approx 0.86\pm0.09$. 
Note that these results are averaged over $r\in[0;1]$ and that $\lambda_2$ increases with $r$ (Fig.\ \ref{fig:redundancy}), 
so that for a properly chosen $r$, the algebraic connectivity can be as low as in the data.\\

Besides for the average shortest path length and the algebraic connectivity, we observed the remaining measures being independent of the system size.\\

Alongside of the Monte-Carlo simulations, for every realization we tested the hypothesis 
that the distribution of degrees larger than four is a geometric distribution with the parameter (i) $\gamma=\hat\gamma$ or (ii) $\gamma=\hat\gamma'$, respectively.
As different goodness-of-fit tests for the geometric distribution perform quite differently \cite{Bracquemond2002},
we chose the generalized Smirnov-transformed Anderson-Darling test $\hat A^2_{GST}$ proposed in \cite{Bracquemond2002},
at a nominal significance level of 5 pct. 
Indeed we find that the test rejects the hypothesis in 5.3 [4.8] pct. ($\hat\gamma$ [$\hat\gamma'$]) of the cases and accepts it in the remaining cases.

\section{Discussion}\label{sec:discussion}
To summarize, we proposed a new growth model to create spatially embedded random networks, that are especially suited to fit the needs of 
modeling power grids. However as indicated in the introduction, this concept might be as well adopted for other infrastructure networks with 
 a similar construction process. By that we refer to the at least two stages (initialization and growth) of development as implemented in our model.
Further it appears also worthwhile to take the splitting of lines into account, as this helps to explain the observed structure of several real world networks,
that might otherwise not be reproducible, thus it became an important feature within our model, not implemented in earlier studies of the topic.
Another key ingredient, minimum spanning trees, capture an important aspect of the initialization, namely, that in early stages of (physical) network construction
 it is often necessary to develop a wide-spreading grid in shortest time. In the later stages however, the notion of redundancy becomes more important 
 for the construction process as well. To capture this, we use a controllable trade-off between redundancy of a link and its spatial length, which we assume to 
 be an important cost factor.\\
 The overall computational complexity is $O(N^3\log N)$, in the absence of line-splitting this can be improved to $O(N^3)$. This means that the complexity of the algorithm is still sufficient for the modeling of large network ensembles as they are often necessary to assess statistical features of complex networks. Thus we hope the model
  becomes a useful tool to assess the properties of power grids and other infrastructure networks.\\
 We have chosen a set of independent model parameters that control the resulting network properties in a relevant range that we explored using Monte-Carlo simulation of our model. Further we found that the hypothesis of an exponentially decaying tail for large degrees in the degree distribution is well accepted for an ensemble of model realizations, using the generalized Smirnov-transformed Anderson-Darling test. However, while the analytically determined mean degree matches the observed one very well, this is not the case for the slope of the exponential decay, which seems to be overestimated with a constant offset. This needs to be taken into account and corrected in the application of the model, however the underlying reason for that remains still unclear to us.\\
Lastly we want to mention possible directions towards extensions of our model. First of all it would be possible to use a more detailed description level of power grids, 
incorporating for instance multiple edges (the N-1 criterion). Also, as our algorithm is partly based on empirical observations, it might only yield realistic high-voltage transmission grids, as distribution grids may show different structural properties \cite{Pagani2011}. We conjecture, that they might be well-described by a model similar to 
our initialization phase, as the apparent redundancy seems to be rather low.\\
More general however, it is also possible to choose different cost functions, adopting to corresponding research questions. For instance it could include geodesic spatial distance measures, taking specific landscapes, population densities or inaccessible areas into account. 
Our ongoing interest lies also on the study of cascading failure processes \citep{Buldyrev2010,Eriksen2003, Motter2002}, i.e., to analyze
 how the proposed networks -- being compared to the standard set of network models -- perform under random failures of or intentional attacks on the infrastructure.


\section*{Appendix A}

The approximate shape of the degree distribution for large $N-N_0$ can be estimated
using a similar approach as in \cite{Callaway2001} for the case of a not spatially embedded random growth model.
For this, we note that the probability of being selected as node $i'$ in step G4 is exactly $q/N$.

If the probability of being selected as node $j$, $\ell$, or $\ell'$ were also the same for all nodes,
the degree distribution for large $N$ could be derived as follows.
Let $d_k(N)$ be the expected number of nodes with degree $k$ after adding $N-N_0$ nodes.
When adding a node via steps G1--G4, it will get degree one or two with probability $1-p$ or $p$, respectively;
on the other hand, the number of degree-$k$ nodes may decrease by at most four 
if such a node is the $j$, $\ell$, $i'$, or $\ell'$ of that step,
which happens with probability $d_k(N)/N$, $pd_k(N)/N$, $qd_k(N)/N$, and $qd_k(N)/N$, respectively,
or may increase by at most four if the selected node has degree $(k-1)$ instead,
which happens with probability $d_{k-1}(N)/N$, $pd_{k-1}(N)/N$, $qd_{k-1}(N)/N$, and $qd_{k-1}(N)/N$, respectively;
when adding a node via step G5, it gets degree two and all other nodes retain their degrees.
Hence the expected number of degree-$k$ nodes evolves as this: 
\begin{align}
	d_1(N+1) &= d_1(N) + (1-s)[(1-p) - (1+p+2q)d_1(N)/N],\\
	d_2(N+1) &= d_2(N) + (1-s)[p + (1+p+2q)(d_1(N)-d_2(N))/N] + s,\\
	d_k(N+1) &= d_k(N) + (1-s)(1+p+2q)(d_{k-1}(N)-d_k(N))/N\quad\text{for~}k>2.\label{eqn:dk}
\end{align}
Plugging the ansatz $d_k(N)\sim Np_k$ into these equations, we get
\begin{align}
	(N+1)p_1 &= Np_1 + (1-s)[(1-p) - (1+p+2q)p_1],\\
	(N+1)p_2 &= Np_2 + (1-s)[p + (1+p+2q)(p_1-p_2)] + s,\\
	(N+1)p_k &= Np_k + (1-s)(1+p+2q)(p_{k-1}-p_k)\quad\text{for~}k>2,
\end{align}
which solves as
\begin{align}
	p_1 &= \frac{(1-s)(1-p)}{1 + (1-s)(1+p+2q)},\label{eqn:p1}\\
	p_2 &= \frac{(1-s)(p + (1+p+2q)p_1) + s}{1+(1-s)(1+p+2q)},\label{eqn:p2}\\
	p_k &= \frac{(1-s)(1+p+2q)}{1+(1-s)(1+p+2q)}p_{k-1} =: \eta p_{k-1}\quad\text{for~}k>2,
\end{align}
hence $p_k$ decays exponentially as $p_k\sim\eta^k = e^{-k/\gamma}$ with  
\begin{align}
	\gamma &= -\frac{1}{\log\eta} = \frac{1}{\log\frac{1+(1-s)(1+p+2q)}{(1-s)(1+p+2q)}} 
		\in \left]0;\frac{1}{\log\frac{5}{4}}\right] \approx ]0;4.48].\label{eqn:gamma}
\end{align}
The decumulative distribution function $P_k=\sum_{k'\ge k}p_k$ then also decays at the same rate.
Note that $\gamma$ increases with both $p$ and $q$, as does $\kappa$,
but there is no one-to-one relationship between $\gamma$ and $\kappa$; 
rather, both $\kappa$ and $\gamma$ can be adjusted via $p,q$ in partial independence.

In reality, however, the probabilities of being selected as node $j$, $\ell$, or $\ell'$ 
may differ between nodes for several reasons,
which explains why the actually observed values of $\gamma$ differ somewhat from the above-derived theoretical value.

\newpage
\section*{Appendix B}

\begin{table}[hpb]
\caption{Data set for the power grids of ENTSO-E member states. Listed are network size $N$, the number of edges $M$, mean degree $\bar k$, the global clustering coefficient ``cc'', the average shortest path length ``aspl'' and the algebraic connectivity $\lambda_2$.}
\centering
\label{tab:ucte}       
\begin{tabular}{lllllll}
\hline\noalign{\smallskip}
member state & $N$ & $M$ & $\bar k$ & cc & aspl & $\lambda_2$\\
\noalign{\smallskip}\hline\noalign{\smallskip}
France                & 3788 & 4907 & 2.591 & 0.059 & 12.260 & 0.003 \\
United Kingdom        & 1712 & 2073 & 2.422 & 0.040 & 15.238 & 0.003 \\
Spain                 & 1512 & 1955 & 2.586 & 0.038 & 11.490 & 0.004 \\
Italy                 & 911  & 1123 & 2.465 & 0.054 & 16.179 & 0.004 \\
Germany               & 820  & 1029 & 2.510 & 0.051 & 15.409 & 0.004 \\
Finland               & 621  & 757  & 2.438 & 0.020 & 10.730 & 0.009 \\
Belgium               & 579  & 778  & 2.687 & 0.050 & 9.503  & 0.011 \\
Austria               & 538  & 632  & 2.349 & 0.025 & 12.127 & 0.002 \\
Hungary               & 528  & 903  & 3.420 & 0.111 & 6.777  & 0.023 \\
Greece                & 436  & 559  & 2.564 & 0.057 & 9.317  & 0.008 \\
Netherlands           & 314  & 392  & 2.497 & 0.080 & 8.033  & 0.009 \\
Sweden                & 265  & 320  & 2.415 & 0.050 & 8.824  & 0.017 \\
Switzerland           & 253  & 320  & 2.530 & 0.055 & 8.737  & 0.014 \\
Moldova               & 223  & 255  & 2.287 & 0.022 & 9.456  & 0.009 \\
Portugal              & 213  & 295  & 2.770 & 0.095 & 6.944  & 0.018 \\
Lithuania             & 202  & 242  & 2.396 & 0.023 & 7.747  & 0.013 \\
Ireland               & 181  & 254  & 2.807 & 0.104 & 5.659  & 0.041 \\
Bosnia \& Herzegovina & 181  & 244  & 2.696 & 0.061 & 5.743  & 0.042 \\
Estonia               & 178  & 233  & 2.618 & 0.083 & 6.509  & 0.025 \\
Norway                & 170  & 209  & 2.459 & 0.016 & 11.752 & 0.003 \\
Croatia               & 165  & 210  & 2.545 & 0.046 & 6.302  & 0.023 \\
Latvia                & 157  & 198  & 2.522 & 0.066 & 6.521  & 0.018 \\
Ukraine               & 152  & 215  & 2.829 & 0.094 & 6.331  & 0.032 \\
Poland                & 144  & 191  & 2.653 & 0.064 & 6.506  & 0.043 \\
Slovenia              & 120  & 155  & 2.583 & 0.147 & 5.615  & 0.032 \\
Denmark               & 118  & 152  & 2.576 & 0.085 & 5.860  & 0.025 \\
Romania               & 102  & 129  & 2.529 & 0.063 & 5.363  & 0.059 \\
Serbia \& Montenegro  & 99   & 124  & 2.505 & 0.117 & 5.701  & 0.030 \\
Albania               & 75   & 89   & 2.373 & 0.018 & 5.564  & 0.043 \\
Luxembourg            & 69   & 81   & 2.348 & 0.004 & 4.858  & 0.046 \\
Macedonia             & 68   & 85   & 2.500 & 0.109 & 4.446  & 0.084 \\
Czech Republic        & 66   & 84   & 2.545 & 0.047 & 4.386  & 0.109 \\
Cyprus                & 60   & 66   & 2.200 & 0.012 & 5.649  & 0.029 \\
Turkey                & 59   & 82   & 2.780 & 0.121 & 5.115  & 0.075 \\
Slovakia              & 45   & 53   & 2.356 & 0.080 & 4.787  & 0.084 \\
Bulgaria              & 44   & 55   & 2.500 & 0.046 & 3.940  & 0.152 \\
Belarus               & 43   & 60   & 2.791 & 0.065 & 3.776  & 0.157 \\
Malta                 & 21   & 29   & 2.762 & 0.263 & 3.138  & 0.131\\
\noalign{\smallskip}\hline
\end{tabular}
\end{table}

\end{document}